\documentclass[10pt,aps,twocolumn,floatfix,showpacs, superscriptaddress, pra]{revtex4-2}

\usepackage{amssymb,amsmath,amsfonts}
\usepackage{graphicx,graphics}
\usepackage{epsfig}
\usepackage{epstopdf}
\usepackage{amsfonts}
\usepackage{bm,bbm}
\usepackage{amssymb,amsmath,amsfonts}
\usepackage{mathrsfs}
\usepackage{calc}
\usepackage{epsfig}
\usepackage{float}
\usepackage{xcolor}
\usepackage{epstopdf}
\usepackage{bm,amssymb,amsmath}
\usepackage{graphicx,color}
\usepackage{bm}

\usepackage{ulem}
\usepackage{xspace}
\usepackage[colorlinks=true, citecolor=teal, urlcolor=purple,linkcolor=purple]{hyperref}
\usepackage{siunitx} 
\newcommand{\specialcell}[2][c]{%
	\begin{tabular}[#1]{@{}c@{}}#2\end{tabular}}
    
\usepackage{amsmath,physics,bm,float}
\usepackage{soul}
\usepackage{amssymb}
\usepackage{natbib}
\usepackage{upgreek}
\usepackage{mathtools}

\usepackage{comment}

\begin{document}


\title{A particle-resolved rheological study of chirality transfer and odd transport}

\author{R\'{e}mi Goerlich}
\email{remigoerlich@tauex.tau.ac.il}
\thanks{Both authors equally contributed to this work}
\affiliation{Raymond \& Beverly Sackler School of Chemistry, Tel Aviv University, Tel Aviv 6997801, Israel}
\affiliation{Institut f\"ur Theoretische Physik II: Weiche Materie, Heinrich-Heine-Universit\"at D\"usseldorf, D-40225 D\"usseldorf, Germany}
\author{Alexander P.\ Antonov}
\thanks{Both authors equally contributed to this work}
\affiliation{Institut f\"ur Theoretische Physik II: Weiche Materie, Heinrich-Heine-Universit\"at D\"usseldorf, D-40225 D\"usseldorf, Germany}

\author{Kristian St\o{}levik Olsen}
\affiliation{Institut f\"ur Theoretische Physik II: Weiche Materie, Heinrich-Heine-Universit\"at D\"usseldorf, D-40225 D\"usseldorf, Germany}

\author{Lorenzo Caprini}
\affiliation{
		Physics department, University of Rome La Sapienza, P.le Aldo Moro 5, IT-00185, Rome, Italy}

\author{Christian Scholz}
\affiliation{Institut f\"ur Theoretische Physik II: Weiche Materie, Heinrich-Heine-Universit\"at D\"usseldorf, D-40225 D\"usseldorf, Germany}

\author{Hartmut L\"owen}
\affiliation{Institut f\"ur Theoretische Physik II: Weiche Materie, Heinrich-Heine-Universit\"at D\"usseldorf, D-40225 D\"usseldorf, Germany}

\author{Yael Roichman}
\email{roichman@tauex.tau.ac.il}
\affiliation{Raymond \& Beverly Sackler School of Chemistry, Tel Aviv University, Tel Aviv 6997801, Israel}
\affiliation{Raymond \& Beverly Sackler School of Physics and astronomy, Tel Aviv University, Tel Aviv 6997801, Israel}

\date{\today}

\begin{abstract}
Chirality, or the breaking of mirror symmetry, appears across all scales in nature, from molecular conformations to the dynamics of bacterial collectives. Environments composed of such symmetry-breaking constituents can give rise to emergent physical phenomena, particularly in the transport and response of embedded tracers.
Yet it remains unclear how chiral environments influence such tracers and through which microscopic mechanisms anomalous responses emerge.
Here, we present a particle-resolved study of these systems, demonstrating chirality transfer and odd transport of an object embedded in a chiral active bath.
In a rheological experiment, a symmetric passive tracer is driven through collisions with the particles of a non-equilibrium chiral bath.
Combining table-top experiments, many-body simulations, and a reduced coarse-grained theory, we demonstrate that local collisions transfer chiral active dynamics to the tracer, which displays circular trajectories.
We show that the same mechanism gives rise to a systematic transverse drift under a constant pulling force.
Crucially, we identify nonlinear friction as an essential factor that rectifies these transferred chiral active fluctuations into a macroscopic odd response.
Our results reveal a microscopic mechanism for odd transport in chiral active matter and provide general insights into transverse transport in driven non-equilibrium systems.
\end{abstract}

\maketitle

\section*{Introduction}

\begin{figure*}[t!]
	\centerline{\includegraphics[width=1\linewidth]{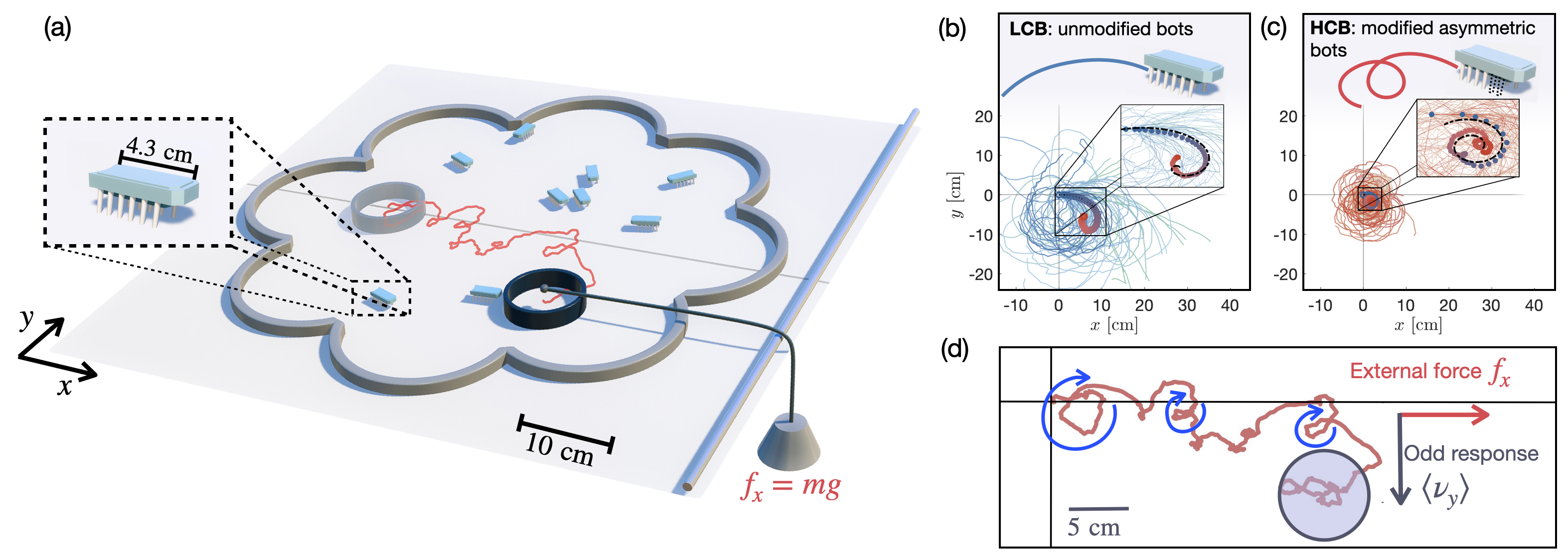}}
	\caption{\textbf{(a) Schematic representation of the experimental setup}.
    A passive tracer (hollow 3D-printed polylactic acid cylinder) is immersed in a 2D bath of self-propelled chiral particles (bristle-bot, shown in the zoomed inset) enclosed in a $R_{\rm A} = 20$ cm-radius arena, whose flower-shaped edge is designed to reduce boundary accumulation. A mass $m$ is attached to the passive tracer and hung outside the arena, applying a constant external force $f_x$ kept aligned with the $x$-axis. The tracer is set into motion by collisions with the bath particles, and undergoes a diffusing trajectory shown in red from an initial tracer position indicated in light color.
    \textbf{(b,c) Trajectories of the environmental chiral active particles}. 65 independent trajectories (shades of blue) of unmodified bots (6 different bots where used), initiated with constant position and orientation (towards positive $x$). The ensemble of trajectories shows a clockwise chirality as confirmed by the mean trajectory obtained through a fit using a logarithmic spiral (black dot-dashed line in the inset of each panel).
    (b) and (c) show the same experiment for unmodified (lowly chiral bath, LCB) and modified (highly chiral bath, HCB) asymmetric bots, respectively. 
    \textbf{(d) Chirality transfer and Hall effect (odd response) of the passive tracer}. We show one example trajectory of the passive tracer immersed in the HCB, showcasing both chirality transfer (loops highlighted with blue arrows) and a Hall effect or odd response (net displacement in the direction normal to the exerted force).}
    \label{fig:Schema}
\end{figure*}

In the classical Hall effect~\cite{nagaosa2010anomalous}, first discovered by Edwin Hall in 1879~\cite{hall1879new}, charge carriers acquire a transverse velocity in response to an applied force, resulting in a current orthogonal to the imposed electric field.
This effect arises from the presence of a magnetic field and constitutes an early example of an \textit{odd transport}.
More generally, odd transport is characteristic for systems that break both parity and time‑reversal symmetry, producing fluxes orthogonal to an applied force.
It appears in diverse systems such as plasmas~\cite{glattli1985dynamical}, gases \cite{hulsman1970transverse}, and magnetized soft matter~\cite{cao2023memory}.
Over the past few years, the concept has gained renewed interest in non‑equilibrium physics, where chirality and active driving such as self-propulsion are susceptible to generate Hall‑like responses even without magnetic fields \cite{poggioli_odd_2023, hargus_odd_2021}.

Active matter, a class of intrinsically nonequilibrium systems, provides an ideal platform for exploring emergent odd phenomena.
It consists of objects that locally convert energy into directed motion thereby breaking time-reversal symmetry at the level of individual trajectories~\cite{elgeti2015physics, marchetti2013hydrodynamics, bechinger2016active}.
When activity is chiral~\cite{lowen2016chirality,liebchen2022chiral}, particles undergo persistent circular motion, thereby also breaking the mirror symmetry at the microscopic level~\cite{van2008dynamics, siebers_collective_2024, caprini_self-reverting_2024, pisegna_spinning_2025, kalz_field_2024}.
Such models are particularly relevant since chirality is widespread in active systems including living organisms ranging from the microscopic \cite{lauga2006swimming,xu2007polarity,huang2021circular,tan_odd_2022} to the macroscopic level \cite{Thomas1993, Obst1996, Souman2009, gu2025emergence}, and engineered systems, including vibrated robots~\cite{barois2020sorting,carrillo-mora_depinning_2025, caprini2025spontaneous} and granular spinners~\cite{soni2019odd, scholz2018rotating, arora2021emergent, lopez2022chirality}.
Importantly, a single active chiral particle in a Stokesian fluid is not sufficient to display odd transport, as its prescribed circular motion does not interfere with the response to an external force \cite{hargus_odd_2021}.

By contrast, a genuine Hall-like mobility can arise in a collective system: a passive tracer immersed in a nonequilibrium chiral bath, where any transverse motion must be generated through interactions with the environment \cite{reichhardt_active_2019, hargus2025odd}.
In that case, the passive tracer may acquire an antisymmetric component in its mobility tensor ~\cite{poggioli_odd_2023}, leading to a spontaneous Hall response. Parity-odd transport coefficients, including odd viscosity and odd diffusivity, have indeed been predicted and analyzed theoretically~\cite{banerjee_odd_2017, reichhardt_active_2019, hargus_odd_2021, lou_odd_2022, kalz2026reversal, hargus2025odd}.
Determining how microscopic chirality and activity is transmitted to a tracer through interactions, and how they reshape the tracer’s effective mobility in a realistic situation is essential for uncovering the mechanisms behind emergent odd transport, and for establishing principles to control motion in such systems~\cite{bowick2022symmetry, mecke_emergent_2024}.
Furthermore, whereas most theoretical studies to date assume a tracer solely interacting with a bath, real active environment typically involve strong interactions and frictional forces~\cite{vega2022diffusive,lopez2022chirality, tan_odd_2022, yang_robust_2020, decurtis_rigid_2025}.

Here, we demonstrate experimentally and numerically that a symmetric passive tracer immersed in a chiral active bath acquires chirality and an odd mobility, manifesting as a spontaneous Hall response (Fig.~\ref{fig:Schema}).
Using a macroscopic rheological platform with direct access to trajectories and collisions, we resolve the mechanism responsible for this behavior.
We show that the combination of chirality and self-propulsion in the bath particles leads to orbital-like collisions with the tracer \cite{sharma_active_2021} with a well-defined handedness, in contrast with purely torque-driven spinner bath particles \cite{reichhardt_active_2019, lopez2022chirality}.
While these interactions induce chirality, in the form of local rotational dynamics of the tracer, applying an external force breaks the spatial symmetry of these trajectories. 
This asymmetry rectifies the local orbital collision statistics into an effective transverse force normal to the external driving direction, generating the odd response.
Crucially, the magnitude of the resulting Hall effect depends sensitively on a non-equilibrium interplay between tracer size, bath chirality, and the substrate's nonlinear frictional regime.
By linking collision-induced chirality transfer to a macroscopic transverse drift, our results provide a minimal physical route to spontaneous Hall-like transport in active matter and identify its mechanical origin.

\begin{figure*}[t!]
    \centerline{\includegraphics[width=1\linewidth]{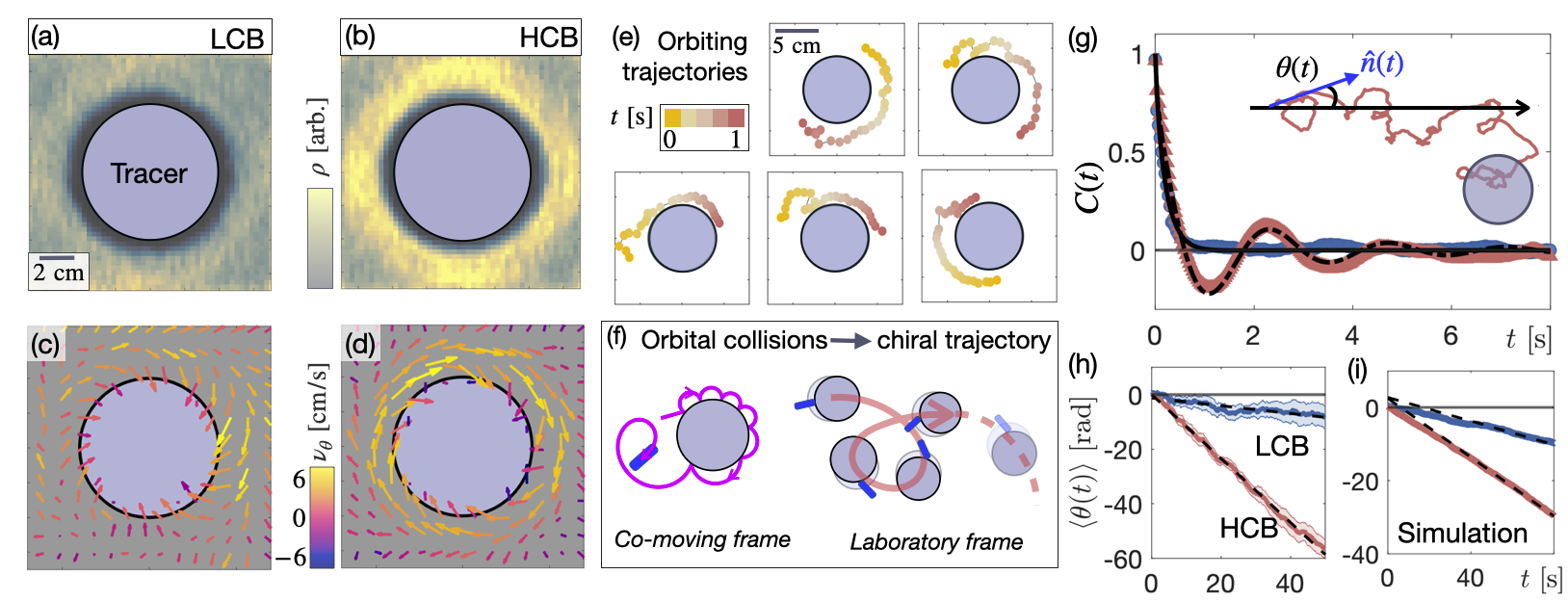}}
	\caption{\textbf{Dynamics of the tracer, from local interactions to chirality transfer.}
    (a)-(b) Local relative density of bots around the tracer in the LCB (a) and HCB (b), measured in a co-moving frame centered on the diffusing passive tracer.
    (c)-(d) Local mean velocity flow around the tracer [color coding the projection $\langle \nu_\theta\rangle$ along the azimuthal direction $\theta$].
    (e) Short illustrative snapshots of orbital trajectories (1 second) measured in the co-moving frame.
    (f) Proposed mechanism at play: orbital radius of bot comparable with tracer radius, $r_c \sim R_{\rm T}$, the bath particle has a significant persistence length and enter in an orbiting and tapping regime, with time-ordered clockwise collisions. This leads to a chiral motion of the tracer.
    (g) Autocorrelation function of the direction of motion of the tracer in the laboratory frame ($\hat n = \boldsymbol{v}/|\boldsymbol{v}|$ where $\boldsymbol{v}$ is the finite-difference velocity of the tracer over a short time-increment $dt = 0.3$ s). The black lines are fits using a cosine with a stretched exponential envelope, $e^{-(t/\tau_0)^{1/2}} \cos(\omega_{\rm T} t)$, where $\tau_0$ and $\omega_{\rm T}$ are two fitting parameters. The inset shows a single tracer trajectory lasting 50 seconds.
    (h)-(i) Mean angular drift of the direction of motion of the tracer in LCB (blue) and HCB (red) measured experimentally (h)
    and numerically (i).}
	\label{fig:Chirality}
\end{figure*}

Although the mechanism originates from collective tracer-bath interactions, we show that a minimal model reproduces the main experimental observations at the single-particle level.
This identifies the key ingredients required for odd transport as the combination of chiral active dynamics and nonlinear friction.
Because the magnitude of the response depends on tracer size, the mechanism also enables selective transport~\cite{poggioli_odd_2023}.
We exploit this effect to demonstrate size-dependent sorting within a chiral active fluid, illustrating how interaction-induced chirality can be harnessed for controlled transport in active environments.

\section*{Results}

\subsection*{Experimental chiral active granular bath}

We experimentally design a chiral bath composed of active particles with tunable chirality, providing a well-controlled framework that captures key features of chiral active matter. By characterizing and modifying individual particle properties, we can tune their interactions with the tracer and the resulting collective behavior, thereby establishing quantitative control over the system.

The active chiral bath is composed of bristle-bots, centimeter-sized self-propelled granular particles \cite{dauchot2019dynamics, chor2023many, boriskovsky2024}, referred to as bots throughout this paper. Their activity is driven by a 3D oscillation powered by batteries, which is converted into forward motion via six bristles on each side of the particle (see \textit{Methods} for details). These bots exhibit a degree of chirality \cite{barois2020sorting} due to an internal body asymmetry, resulting in the circular-like trajectories characteristic of chiral active matter.
To increase the chirality of the bots, we remove two bristles on the right side of the particles, introducing an asymmetry in their propulsion mechanism. These two classes of particles form active baths with distinct chiralities: the unmodified particles constitute a low chirality bath (LCB, Fig.~\ref{fig:Schema}~(b)), while the modified asymmetric particles form a high chirality bath (HCB, Fig.~\ref{fig:Schema}~(c)).
The difference in chirality is confirmed by their stochastic circular trajectories, as well as by their averaged trajectories starting from a fixed initial orientation, which give rise to logarithmic spirals \cite{shenoy2007kinematic, van2008dynamics, kummel2013circular, hargus_odd_2021}. Their mean radius of curvature $r_c$ decreases from the LCB to the HCB, reflecting the increase in chirality.
As shown in \textit{Methods}, the free bots are well described by the chiral active Brownian particles model \cite{Kiechl2026}.
In contrast to chiral fluids composed of purely rotating spinners \cite{vega2022diffusive, lou_odd_2022}, these particles maintain circular-like trajectories with a finite mean radius due to the presence of self-propulsion \cite{caprini2025spontaneous}.

To create a chiral active bath, we enclose $N=20$ chiral active bots within an arena of radius $R_{\rm A}=20$ cm.
Both the floor and edges of the arena are made of poly-vinyl and its design follows a circular octagonal shape, which reduces boundary accumulation. 
Compared to the single bot dynamics studied above (Fig.~\ref{fig:Schema} (b)-(c)), the collective dynamics of the active particles are modified by (i) boundary effects and (ii) collisions.
The properties of the active particles in the bath are characterized at the individual and collective levels in \textit{Methods} (see Fig.~\ref{fig:BathChirality}).

\subsection*{Chirality transfer from bath to the passive tracer}

A passive tracer with radius $R_{\rm T} = 4~\rm{cm}$ is immersed in this bath and subjected to a constant external force along the $x$-axis, applied by hanging a light mass (see Fig.~\ref{fig:Schema}(a) and \textit{Methods} for details).
We follow the tracer's position along its diffusive trajectory, set to motion by collisions with the bath particles and biased by the external force.
In Fig.~\ref{fig:Schema}~(d), we show a single recorded tracer's trajectory, revealing both its inherited chirality and the odd response to the external force.
In what follows, we will quantify and explain these two distinct effects.

To understand the impact of bath chirality, we track the motion of bath particles in a co-moving frame of reference along the tracer's trajectory. As seen in Fig.~\ref{fig:Chirality}~(a-d), HCB and LCB interact in distinct ways with the tracer.
Increasing the chirality of the bath particle leads to an accumulation of bots around the tracer (Fig.~\ref{fig:Chirality}(a)-(b)) as well as a strong clockwise circulation along the tracer's edge (Fig.~\ref{fig:Chirality}(c)-(d)).
This behavior arises from the relative size difference between the chiral radius of the trajectory $r_c$, compared to the radius of the tracer $R_{\rm T}$.
When $r_c \lesssim R_{\rm T}$, self-propulsion causes particles to accumulate at the tracer boundary, leading to persistent orbiting trajectories of the bots around the tracer (Fig.~\ref{fig:Chirality}(d)).
In Fig~\ref{fig:Chirality}(e) we show experimental evidence of the individual bot orbiting trajectories that lead to the clockwise flow in the HCB.
Such orbiting interactions are at the origin of chirality transfer, as we show in Fig.~\ref{fig:Chirality}(f).
The mechanism at play is the following, a bath particle undergoes time-ordered repeated short-time recollisions (‘tapping’) with the tracer, effectively sliding clockwise along its surface.
These time-ordered tapping collisions create a handed sequence of impulses that breaks mirror symmetry and induces circular-like motion of the tracer in the lab frame.
In other words, the tracer acquires a degree of chirality.
For larger $r_{\rm c}$, orbiting trajectories are reduced, decreasing the clockwise correlations of tapping collisions that lead to chirality transfer.

The resulting transfer of chirality on the tracer's dynamics can be experimentally quantified by measuring the autocorrelation function $C(t)$ of its direction of motion $\hat n$ (Fig.~\ref{fig:Chirality}~(g)).
When the tracer is immersed in the LCB, $C(t)$ quickly decays to 0, without visible oscillation.
However, the mean relative angle $\langle \theta(t)\rangle$ (between $\hat n$ and the $x$-axis) demonstrates that the direction of motion of the tracer is constantly drifting (Fig.~\ref{fig:Chirality} (h)), which reveals a weak chiral dynamics.
When the tracer is immersed in the HCB, $C(t)$ oscillates, as the correlation function of a chiral active particle, yet with a stretched exponential envelope $C(t) = e^{-\sqrt{t/\tau_0}} \cos{(\omega_{\rm T} t)}$, characterized by the fitting parameter $\tau_0$ and the induced chirality $\omega_{\rm T}$.
Similar stretched exponential correlations have been reported for Brownian motion in power-law viscoelastic baths \cite{quevedo2025active}. 
The mean angular drift $\langle \theta(t)\rangle$ shown in Fig.~\ref{fig:Chirality}(h) that leads to estimations of the chirality acquired by the tracer $\omega_{\rm T} = 0.14 \pm 0.08~\rm{rad/s}$ in the LCB and $\omega_{\rm T} = 1.17 \pm 0.06~\rm{rad/s}$ in the HCB.

These results demonstrate that the passive tracer is activated by collisions with the bath particles.
It inherits persistent motion, but does not follow the same exponential correlation as the bath particles that could have been expected from a direct transfer of the bath properties.
Beyond mere activation, the tracer also becomes chiral, following trajectories with a well-defined average handedness.
The contrasting tracer's properties in LCB versus HCB (Fig.~\ref{fig:Chirality}~(g)) and the associated co-moving fields (Fig.~\ref{fig:Chirality}~(c)-(d)) indicate that the chirality transfer arises from a change in local interaction, in the absence of any intrinsic asymmetry in the tracer.
As discussed in \textit{Methods}, the spinning of the tracer around its own axis can be neglected: chirality transfer only depends on transverse collisions, irrespective of rotational friction.

\begin{figure}[t!]
	\centerline{\includegraphics[width=1\linewidth]{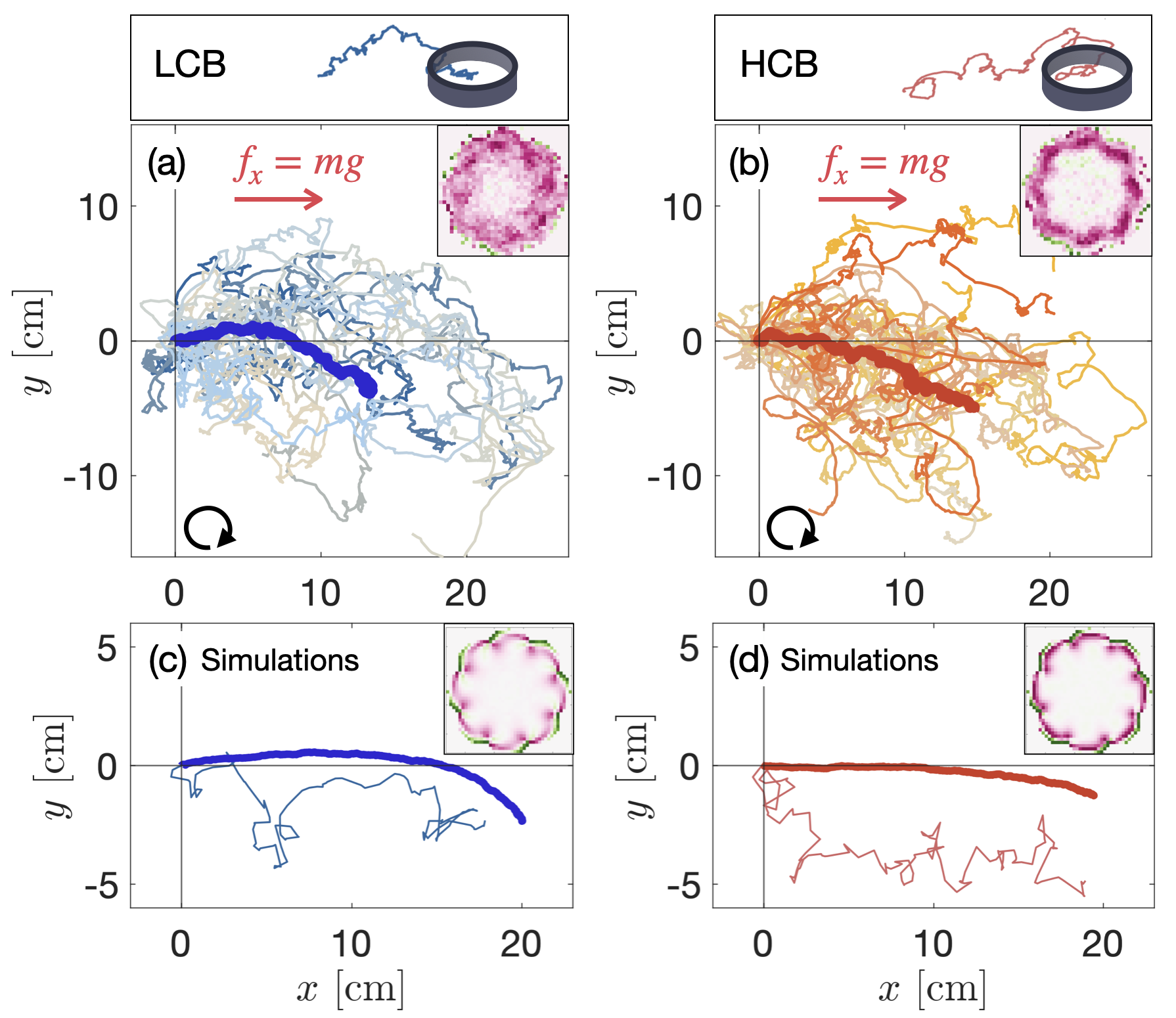}}
	\centerline{\includegraphics[width=1\linewidth]{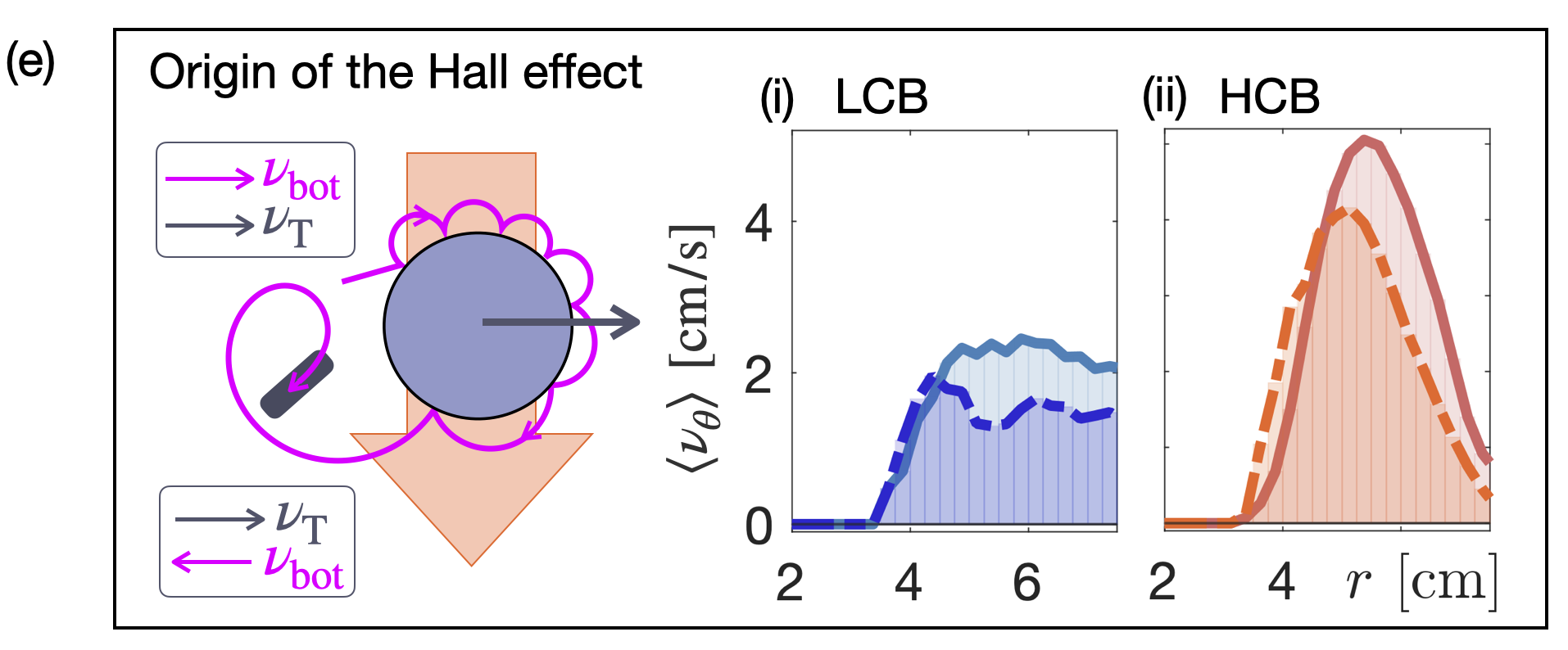}}
	\caption{\textbf{Spontaneous Hall effect (odd transport) of a passive tracer in a chiral bath under an external force}.
    (a)-(b) Recorded trajectories of the passive tracer in the chiral active bath together with the ensemble averaged trajectory (thick lines) both in the LCB (b, blue lines) and in the HCB (c, red lines).
   The transverse response of the passive tracer is revealed by its mean drift towards negative $y$, orthogonal to the applied force.
    In the inset, we reproduce the azimuthal velocity field given in Fig.~\ref{fig:BathChirality} (see \textit{Methods}, ruling out the effect of global edge current in the HCB since in that case, the fluid in the bulk is static.
    (c)-(d) Mean trajectory measured on an ensemble of 5000 numerical simulations with the same boundary conditions as in the experiments. In the inset, we show the respective azimuthal velocity field.
    (e) The orbital interaction regime leads to an odd response to the external force $f_x$. This arises from the different relative velocities on the two sides of the tracer along the vertical $y$-axis, resulting in asymmetric collision statistics.
The orange arrow illustrates the odd response.
    Inset: for the LCB (i) and HCB (ii) cases, we show the experimental evidence of the change in relative velocity above and below the tracer, breaking symmetry along the $y$-axis. The dashed lines correspond to the radial distribution of orbital velocity restricted to positive $y$ (averaged over the angle $\theta \in [0, \pi[$); the solid line to negative $y$ (average over $\theta \in [\pi, 2\pi]$).
    }
	\label{fig:Traj}
\end{figure}

The experimental results presented here are complemented by simulations of chiral active Brownian particles interacting via purely transversal collisions with a passive tracer in a confining arena (see \textit{Methods} for details). Our numerical study is qualitatively consistent with the experimental findings for both the low and high chirality cases. In particular, simulations reproduce the mean angular drift $\langle \theta(t)\rangle$ of the passive tracer observed experimentally (compare Fig.~\ref{fig:Chirality}~(h) with Fig.~\ref{fig:Chirality}~(i)).
This result confirms that the chirality of the bath particles is the key factor driving chirality transfer to the tracer.


\subsection*{Spontaneous Hall effect under constant forcing}

A tracer immersed in an active bath and subjected to an external force $f_x$ constitutes an active microrheology experiment, as the tracer’s response depends on the properties of the surrounding medium. Our experiment shows that the tracer responds transversely to the applied force thereby showing a macroscopic Hall effect.

\begin{figure*}[t!]
	\centerline{\includegraphics[width=0.95\linewidth]{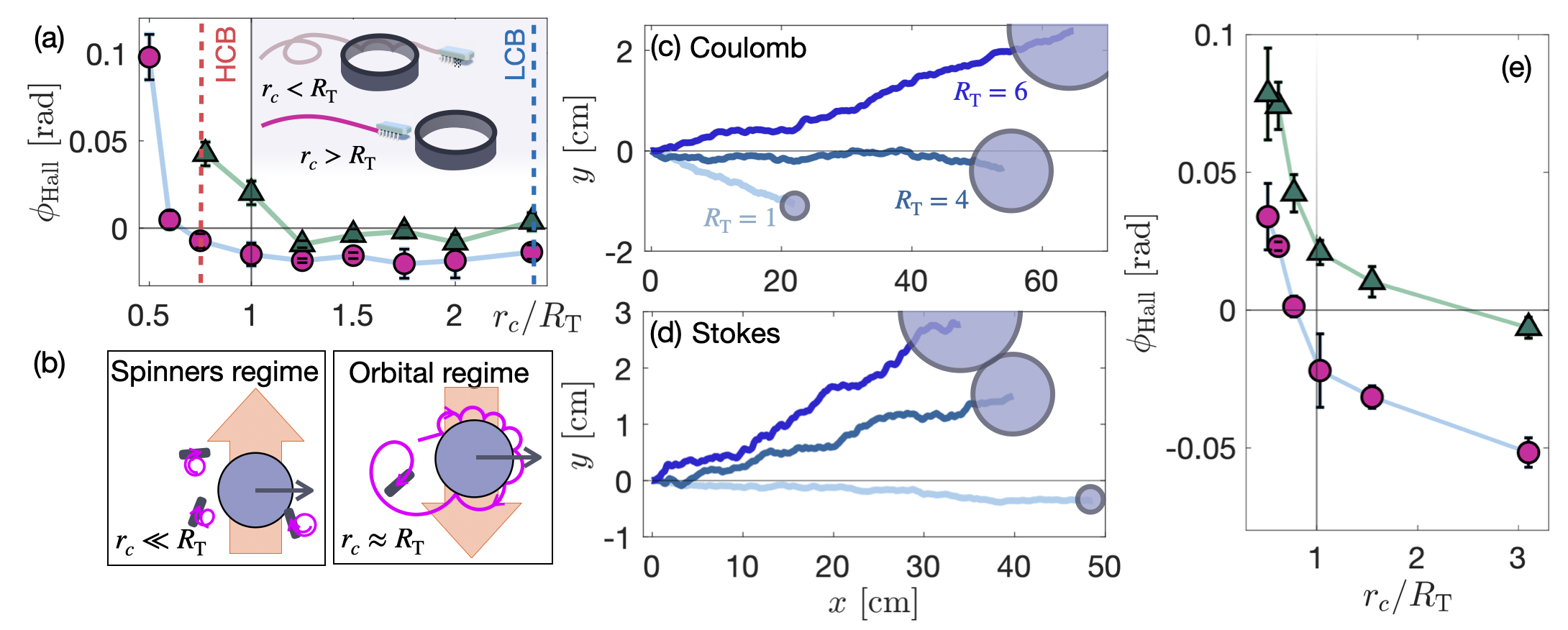}}
	\caption{\textbf{(a,b) Lengthscale-induced reversal of Hall effect, numerical simulations}.
    (a) Hall angle $\phi_{\rm Hall}$ measured in many-body numerical simulations with periodic boundary conditions for varying chirality $r_c$ and a tracer of fixed radius $R_{\rm T}$ with Stokes friction (green triangles) and dry (Coulomb) friction (pink circles), as in the experiment. The vertical dashed line underline the values of $r_c$ of the LCB and HCB (blue and red respectively) used in the experiment. The data are shown with a 3-sigma errors estimated with a bootstrapping method in the linear fitting of the odd response trajectories.
    (b) Illustrations of two regimes of chiral bath particles: (left panel) for very high chirality ($r_c \ll R_{\rm T}$) the bath is effectively composed of spinners \cite{reichhardt_active_2019} that leads to a negative odd transport in contrast to the orbital regime of our experiments (right panel).   
    \textbf{(c-e) Size-dependent spatial sorting of non-chiral objects by a chiral active fluid.}
    Mean trajectories of tracers with varying radii $R_{\rm T}$ in a chiral active fluid ($r_c = 3~\si{cm}$), obtained by averaging 5000 numerical realizations.
    (c) Tracers with $R_{\rm T} = [2, 4, 6]~\si{cm}$ under Coulomb friction.
    (d) Same tracers under Stokesian friction.
    (e) Hall angle as a function of $R_{\rm T} / r_{\rm c}$ for Coulomb (pink circles) and Stokes (green triangles) frictions.
    According to the proposed mechanism for the odd response (see Fig.~\ref{fig:LengthScales} (d)), smaller tracers exhibit a Hall effect, while larger tracers display an anti-Hall effect. Additionally, Stokes friction tends to enhance the anti-Hall behavior compared to the Coulomb friction.}
	\label{fig:LengthScales}
\end{figure*}

In Fig.~\ref{fig:Traj}~(a),b, we present individual tracer trajectories alongside their ensemble averages (thick blue and red lines) within the low-chiral (LCB) and high-chiral (HCB) baths, respectively. While the expected longitudinal drift parallel to the forcing direction ($x$-axis) is observed in both cases, the tracer also develops a distinct, net transverse velocity along the negative $y$-direction. This motion provides a striking macroscopic signature of a spontaneous Hall effect, characterized by non-zero off-diagonal components $\mu_{\text{odd}}$ in the mobility tensor $\boldsymbol{\mu}$ defined via $\mathbf{v} = \boldsymbol{\mu}\mathbf{F}$ \cite{poggioli_odd_2023, hargus2025odd}. Crucially, the quiescent bulk of the HCB, where $\langle \nu_\theta \rangle \approx 0$ in the central region, demonstrates that this transverse transport cannot be attributed to passive advection by global, boundary-driven currents. This is further supported by the fact that the transverse motion persists even when edge-induced flows are entirely suppressed in the bulk (see Fig.~\ref{fig:Traj}~(a)-d insets, and \textit{Methods}), cementing the fact that the measured odd transport arises from local non-equilibrium interactions between the tracer and the chiral active particles.

To uncover the mechanical origin of this response, we resolve the local tracer-bath kinematics. As shown in the previous section on chirality transfer (Fig.~\ref{fig:Chirality}), in our experiments, the radius of the orbital trajectory is comparable to the tracer radius ($r_{\text{c}} \sim R_{\text{T}}$) and the interaction between the bots and the tracer are orbital \cite{sharma2021active} and exhibits a tapping-like character \cite{caprini_emergent_2024} with multiple recollisions. Furthermore, under an external force $f_x$, the tracer drifts with an average longitudinal velocity $\nu_{\text{T}}$. This forward motion breaks the lateral symmetry along the $y$-axis by inducing an asymmetry in the relative velocity between the moving tracer and the orbiting bots ($\nu_{\text{bot}}$) on opposite sides of the object (the upper and lower half-planes in Fig.~\ref{fig:Traj}e). Consequently, the frequency of tapping collisions becomes imbalanced, generating a net transverse bias in the negative $y$-direction. 

To understand the transverse response, we first examine the local collision geometry.
The orbital regime ($r_{\text{c}} \sim R_{\text{T}}$), where bath particle orbits are comparable to the tracer size, differs fundamentally from the diffuse limit ($r_{\text{c}} \gg R_{\text{T}}$) where chirality vanishes and approaches are uncorrelated. 
In the orbital regime, successive collisions are time-ordered, yielding constructive momentum transfer.
In the presence of the force $f_x$ that breaks the spatial symmetry, we experimentally measure a $2.6\times$ relative acceleration of the bots in the lower-hemisphere of the co-moving frame, leading to a lower collision rate directly driving the macroscopic transverse drift. 
Conversely, in the diffuse limit, collisions will randomize and cancel as a standard random walk, 
which should suppress the odd response. 

Intriguingly, despite their substantially different tracer chiralities ($\omega_{\text{T}}$, Fig.~\ref{fig:Chirality}h), the LCB and HCB regimes yield comparable Hall response magnitudes (Fig.~\ref{fig:Traj}~(b,c). This insensitivity reflects a distinct length-scale dependence between local chirality transfer and macroscopic transport. While local chirality transfer depends strongly on the orbital stability and coherence of tapping collisions, the macroscale Hall angle is controlled by the force-induced collision asymmetry. Although both phenomena are ultimately governed by the geometric ratio $r_{\text{c}}/R_{\text{T}}$, they can follow distinct functional forms.
To test this interpretation beyond the limited range accessible experimentally and map the complete length-scale landscape, we turn to many-body simulations with periodic boundary conditions.
These simulations, presented in Fig.~\ref{fig:LengthScales}~(a), systematically sweep $r_{\text{c}}/R_{\text{T}}$, showing that within the orbital-interaction regime, the Hall angle displays a broad maximum that varies only weakly with $r_{\text{c}}/R_{\text{T}}$. Both experimental baths lie within this broad finite-response region ($r_{\text{c}}/R_{\text{T}} \simeq 2.4$ for the LCB and $r_{\text{c}}/R_{\text{T}} \simeq 0.78$ for the HCB).
In the following section, we explore the full geometric dependence, unveiling a dramatic sign reversal at very high chirality and eventual suppression at very large scales.

\subsection*{Length-scale controlled reversal of odd mobility}

Numerical simulations overcome experimental limitations by enabling a broader exploration of physical parameters and isolating the odd-response mechanism from boundary effects.
By implementing periodic boundary conditions, we can remove the influence of the arena walls and confirm that the odd transport persists independently of boundary effects.
In that case, the tracer follows on average straight lines forming a so-called \textit{Hall angle} $\phi_{\rm Hall}$ with respect to the $x$-axis along which the external force is applied (see \textit{Methods} for details) \cite{reichhardt_active_2019, kalz2026reversal}.
Accordingly, a nonzero Hall angle denotes Hall-like transverse transport, which can be positive or negative, corresponding to positive or negative odd response, respectively \cite{kalz2026reversal}.
In Fig.~\ref{fig:LengthScales}~(a), we show $\phi_{\rm Hall}$ as a function of the ratio $r_c/R_{\rm T}$ between the chirality radius of the bath particles and the tracer's radius.
The presence of a finite Hall angle under periodic boundary conditions confirms that odd mobility arises from the intrinsic odd interactions, rather than from boundary-induced currents.

The Hall angle $\phi_{\rm Hall}$ follows a non-monotonic evolution as a function of the ratio $r_c/R_{\rm T}$, reaching a broad maximum when $r_c \gtrsim R_{\rm T}$.
This confirms, in the absence of boundary effects, that the HCB and LCB parameters probed experimentally led to a similar Hall effect.
It reaches negative values for small $r_c/R_{\rm T}$, \textit{i.e.,} very chiral particles.
Looking back at the mechanism for odd response shown in Fig.~\ref{fig:Traj}(a), we propose the following explanation for its reversal for high chirality.
For a very chiral bath ($r_c \ll R_{\rm T}$) the bath particles behave like diffusive spinners (Fig.~\ref{fig:LengthScales}(b)) while the amplitude of translational and rotational noise relative to chirality prevents orbiting trajectories.
In that case, we recover the same sign of Hall effect as in previous works with a bath of spinners \cite{reichhardt_active_2019}.
In this regime, the tracer gradually moves towards positive $x$ and hence meets the spinners more often as it is moving up towards positive $y$ leading to an anti-Hall effect with respect to our experiments.
As discussed in \textit{Methods}, the rotation of the tracer around its own axis and associated rotational friction can be neglected.


\subsection*{Role of nonlinear friction and selective sorting}

All the experimental and numerical results discussed above correspond to a tracer experiencing Coulomb (dry) friction with the underlying substrate. Unlike the usual Stokes friction, which is proportional to the particle velocity, Coulomb friction exerts a velocity-independent force that opposes the direction of motion with a finite threshold. Here, we employ numerical simulations of active particles to investigate how the type of friction between the tracer and the substrate affects the emergence of odd transport. We find that nonlinear friction strongly enhances the positive transverse response in the regime of our experiments. 

In Fig.~\ref{fig:LengthScales}(a), we compare the Hall angle $\phi_{\rm Hall}$ measured for active Brownian particles subject to either Coulomb (dry) or Stokes friction (see \textit{Methods} for details of the numerical simulations). When the ratio of the circular trajectory radius of the bots to the tracer radius, $r_{\rm c}/R_{\rm T} \ll 1$, is small, an anti-Hall effect (sign reversal of $\phi_{\rm Hall}$) is strong in the Stokes friction case, consistent with previous numerical results~\cite{kalz2026reversal}. However, for $r_{\rm c}/R_{\rm T} \gtrsim 1$, the magnitude of the positive Hall effect is significantly reduced with Stokes friction.
As a result, the transverse response becomes barely visible compared with the dry-friction case. Thus, while Coulomb friction is not strictly required for odd transport, nonlinear friction substantially amplifies the positive odd mobility in a chiral active bath.

As shown above, the Hall angle depends on the interplay between two length scales, $r_{\rm c}$ and $R_{\rm T}$.
This dependence suggests that chiral active baths can sort passive objects by size \cite{poggioli_odd_2023}, a practical challenge in both microfluidic and granular systems~\cite{sajeesh2014particle}. For example, biochemical methods have long sought ways to separate chiral objects from their mirror twins across different scales~\cite{barois2020sorting, tkachenko2014optofluidic, 
Meinhardt2012separation}, typically relying on enantiomer-specific interactions with chiral environments~\cite{genet2022chiral}.
In contrast, the Hall effect mechanism we reveal, rooted in the self-propelled trajectories of the bath particles, couples the bath’s chirality even to non-chiral objects, enabling a novel size-dependent sorting of symmetric objects.
Using the many-body numerical simulations discussed above, we  demonstrate this effect for both Coulomb and Stokesian friction (see Fig.~\ref{fig:LengthScales}(c-e)).
Such a mechanism could be extended to microfluidic sorting using biological or artificial microscopic chiral baths \cite{kummel2013circular, li_robust_2024}.

\subsection*{Minimal model}
\begin{figure}[t!]
	\centerline{\includegraphics[width=1\linewidth]{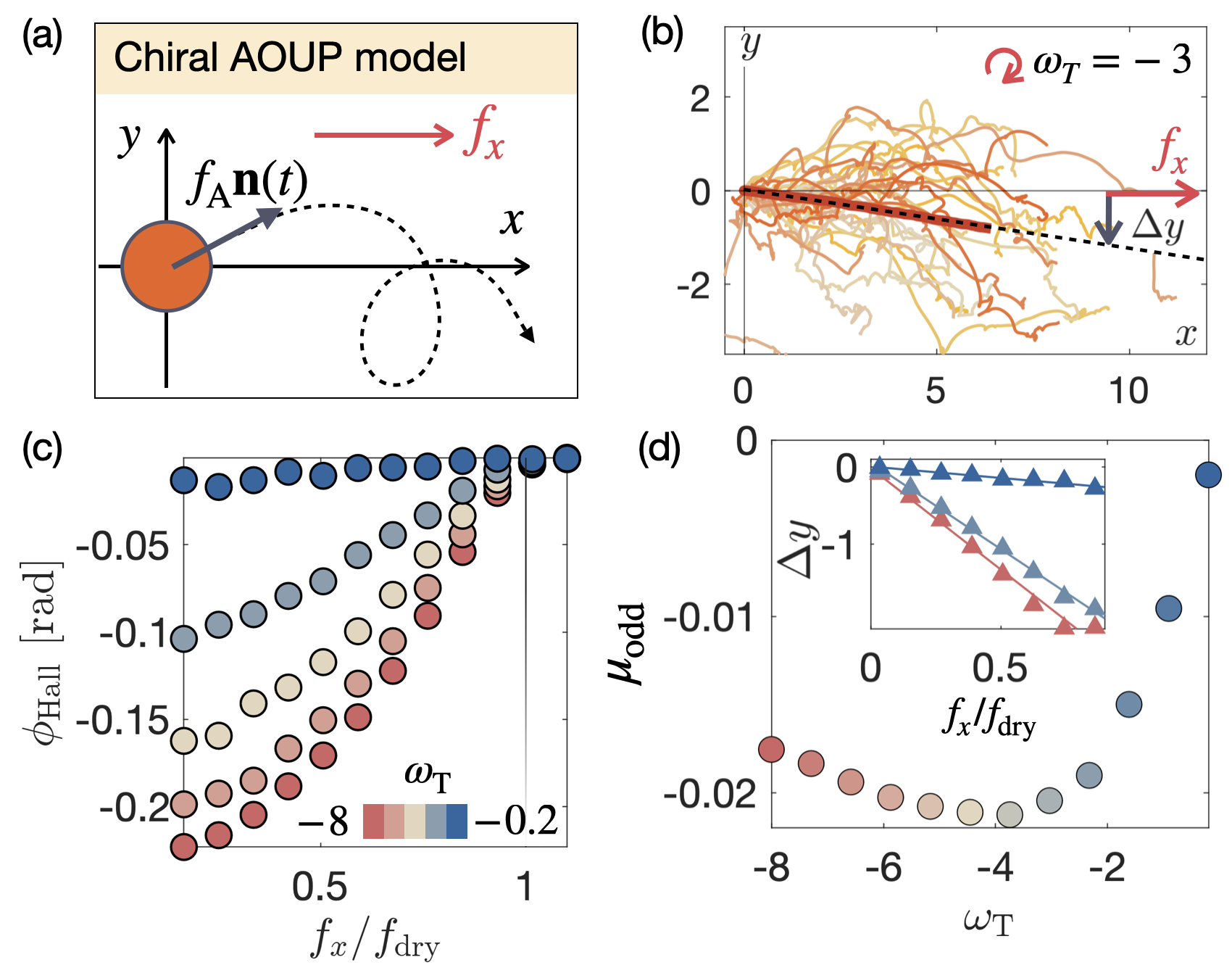}}
	\caption{\textbf{Minimal model of a single chiral active Ornstein-Uhlenbeck particle (chiral AOUP).}
(a) Schematic representation of the minimal model capturing the orbiting regime. The tracer particle is driven by an effective chiral active force $f_{\text{A}}\mathbf{n}(t)$, where $\mathbf{n}(t)$ follows a chiral Ornstein-Uhlenbeck process, and is subjected to a constant external force $f_x$ along the $x$-axis under nonlinear dry friction.
(b) Ensemble of 1000 simulated trajectories generated via the chiral AOUP model [Eq.~\eqref{Eq:CAOUP}] for parameters $\tau_p=0.25$, $f_{\text{A}}=2$, $f_x=1$, and $f_{\text{dry}}=3$, using the Euler-Maruyama integration method with a time step $dt=10^{-2}$ and total duration $t=50$. The ensemble-averaged trajectory (thick red line) demonstrates the spontaneous emergence of an odd response within this reduced model framework.
(c) Hall angle $\phi_{\text{Hall}}$ as a function of the applied force $f_x$, normalized by the dry-friction threshold $f_{\text{dry}}$, for increasing chirality (from blue to red).
(d) Odd mobility coefficient $\mu_{\text{odd}}$ as a function of the particle chirality $\omega_{\text{T}}$. The coefficient $\mu_{\text{odd}}$ is extracted from the linear relationship between the transverse displacement $\Delta y$ and the applied force $f_x$ (inset). In panels (c,d), each data point is computed from the final transverse position $y$ at $t=100$, averaged over 15 independent batches of 1000 trajectories. Unless specified otherwise, parameters match those in panel (b) and are reported in arbitrary units.}
\label{fig:Model}
\end{figure}

To isolate the minimal ingredients required for odd mobility, we consider a coarse-grained single-particle description of the tracer. Because the active bath transfers effective activity and chirality to the embedded object, its motion can be modeled as a chiral active Ornstein-Uhlenbeck particle (chiral AOUP, Fig.~\ref{fig:Model}~(a)) \cite{caprini2019active} under a constant external force:
\begin{subequations}
\label{Eq:CAOUP}
\begin{align}
m_{\text{T}}\frac{\mathrm{d} \bm{\nu}}{\mathrm{d} t} & = -f_{\text{dry}} \hat{\bm{\nu}} + f_x \mathbf{e}_x + f_{\text{A}} \mathbf{n}, \\
\frac{\mathrm{d} \mathbf{n}}{\mathrm{d} t} & = -\frac{\mathbf{n}}{\tau_p} + \omega_{\text{T}}\, \mathbf{n} \times \mathbf{e}_z + \sqrt{\frac{2}{\tau_p}}\boldsymbol{\chi}.
\end{align}
\end{subequations}
Here $m_{\text{T}}$ is the tracer mass, $\bm{\nu}$ is its velocity, $\hat{\bm{\nu}} = \bm{\nu}/|\bm{\nu}|$ is the direction of motion (with $\hat{\bm{\nu}}=0$ for $\bm{\nu}=0$), $f_{\text{dry}}$ is the dry-friction threshold, $f_x$ is the external force along $\mathbf{e}_x$, $f_{\text{A}}$ is the active force amplitude, $\mathbf{n}$ is the active propulsion director, $\tau_p$ is the persistence time, $\omega_{\text{T}}$ is the transferred chirality, and $\boldsymbol{\chi}$ is Gaussian white noise.

This reduced model is not a parameter-free fit of the experiment, but an effective description of the tracer after local tracer-bath collisions have established persistent chiral active dynamics. Its purpose is to isolate the role of nonlinear dry friction. Once effective chiral active dynamics is present, dry friction provides a rectification mechanism that converts these fluctuations into a macroscopic transverse drift. By contrast, in the corresponding single-particle model with linear Stokesian dissipation, chiral active fluctuations yield an \textit{odd diffusivity}~\cite{hargus_odd_2021} but do not generate a steady transverse response to a constant external force.

In Fig.~\ref{fig:Model}~(b), simulated chiral AOUP trajectories qualitatively capture the experimental Hall response (Fig.~\ref{fig:Traj}~(b)), producing a transverse drift toward negative $y$ under clockwise internal chirality. This shows that a single chiral active particle can exhibit an odd mobility when its dissipation is nonlinear, consistent with frameworks in which Coulomb friction breaks detailed balance when coupled to stochastic noise \cite{manacorda2014coulomb, gnoli_brownian_2013}.

Mapping the parametric response reveals that the emergent Hall angle $\phi_{\text{Hall}}$ is maximal at small driving forces and decays to zero as $f_x$ approaches $f_{\text{dry}}$ (Fig.~\ref{fig:Model}~(c)). In this regime, the external force increasingly dominates the dynamics, reducing the relative contribution of transverse chiral fluctuations. Finally, while the transverse displacement scales linearly with small forces, $\nu_y = \mu_{\text{odd}} f_x$ (Fig.~\ref{fig:Model}~(d), inset), the extracted odd mobility $\mu_{\text{odd}}$ varies non-monotonically with the effective chirality $\omega_{\text{T}}$ (Fig.~\ref{fig:Model}~(d)). This non-monotonicity indicates that odd transport is optimized at an intermediate chirality, where the rotation of the active drive is neither too slow to bias the motion nor too fast to average out. The reduced model therefore reinforces the central conclusion that the active Hall effect arises from the coupling between transferred chiral active fluctuations and nonlinear substrate dissipation.


\section*{Discussion}

In this work, we have established a macroscopic rheological framework to investigate how the emergent chirality and transverse response of a driven tracer depend on the non-equilibrium properties of a surrounding chiral active bath.
By resolving local tracer-bath kinematics, our experiments and complementary many-body simulations identify the mechanism driving the odd response.
First, the time-ordered orbital momentum transfers allows cross-scale chirality transfer \cite{pisegna_spinning_2025} from the bath to the tracer.
Second, under the action of an external force, this local chirality transfer leads to a macroscopic odd transport along the transverse direction.

Crucially, our study focuses on a dry active matter system, where both the active bath particles and the passive tracer are governed by nonlinear substrate friction. 
Rather than acting as a non-ideal experimental constraint, this substrate-mediated nonlinear friction emerges as an indispensable physical feature that rectifies local active chiral fluctuations into odd transport.
By contrast, previous theoretical works have shown that, while the broken symmetries of chiral active matter can induce odd diffusivity~\cite{hargus_odd_2021}, linear Stokesian dissipation inherently suppresses transverse response to a driving force.




These findings distinctly separate our mechanism from existing non-equilibrium frameworks such as odd viscosity~\cite{banerjee_odd_2017, fruchart2023odd}, torque-driven spinner fluids~\cite{reichhardt_active_2019}, or Magnus effects~\cite{cao2023memory}. Here, the observed response is generated by local, time-ordered collision statistics intrinsically coupled to nonlinear substrate friction, rather than by global hydrodynamic flows or an intrinsic tracer spin. Furthermore, our minimal chiral active Ornstein-Uhlenbeck particle description demonstrates that nonlinear friction enables even a single-particle model of chiral active motion to exhibit an odd response to an external force.

Our results suggest that chirality-induced effects observed in nature \cite{tan_odd_2022} may be strongly enhanced, or even enabled, by nonlinear dissipation. This insight is directly relevant to active systems governed by nonlinear surface friction, contact-mediated drag, or stick-slip dynamics, ranging from active granular media and crawling cells to soft robotic platforms and social insects performing collective transport \cite{hu2016entangled}. Beyond identifying a minimal mechanism for spontaneous Hall-like transport, these principles offer a general design paradigm for engineering directed motion and mechanical work extraction \cite{behn2017dynamics, bonomo2024sokoban} in synthetic metamaterials and cooperative non-equilibrium systems operating on dissipative substrates.

\acknowledgements
The authors thank Iman Abdoli, Kai Luca Spanheimer, Marco Musacchio, Erik Kalz and Jonas Veenstra for insightful discussions.
R.G. and Y.R. acknowledge support from the European Research Council (ERC) under the European Union’s Horizon 2020 research and innovation program (Grant Agreement No. 101002392). R.G. acknowledges support from the Mark Ratner Institute for Single Molecule Chemistry at Tel Aviv University.
KSO acknowledges support from the Alexander von Humboldt Foundation.

\section*{Methods}

\subsection*{Experimental methods and tracking}

In order to perform active rheology measurement on this system, a tracer in the form of a 3D-printed hollow PLA cylinder of radius $R_T = 4~\rm{cm}$ is subjected to a constant force towards positive $x$, by attaching to it a ball of mass $m_{\rm ball} = 0.87\pm0.01~\rm{g}$ (see Fig.~\ref{fig:Schema}, see Table \ref{Table1} for extensive list of the physical parameters).
The gravitational drift force experienced by the tracer is $f_x = m_{\rm ball} g \approx 8.5~\rm mN$ with $g\approx 9.81 ~\rm{ms^{-2}}$ the gravitational acceleration.
Against the solid surface, the passive tracer experiences a dry friction $\mathbf{f}_{\rm dry} = -f_{\rm dry} \hat{\bm{\nu}},~ \hat{\bm{\nu}} = \bm{\nu}/|\bm{\nu}|,$ with $f_{\rm dry} \approx 0.48 \times m_{\rm T}g = 63.1 ~ \rm{mN}$.
The dry friction largely overcomes the external gravitational force, therefore, in the absence of collision with the bath particles, the tracer is at rest.
The motion of the tracer and particles in the plane is recorded with a camera (Logitech Brio).
Its trajectory is probed via the tracking of four circular stickers, the mean position of the stickers corresponds to the center of the tracer. 
The bots in the laboratory and in the co-moving frame are tracked using a custom-developed motion tracking algorithm, identifying patterns in the differential image of two consecutive frames.
Their trajectories are reconstructed by linking the nearest tracked location between consecutive frames.

\begin{figure*}[ht!]
	\centerline{\includegraphics[width=1\linewidth]{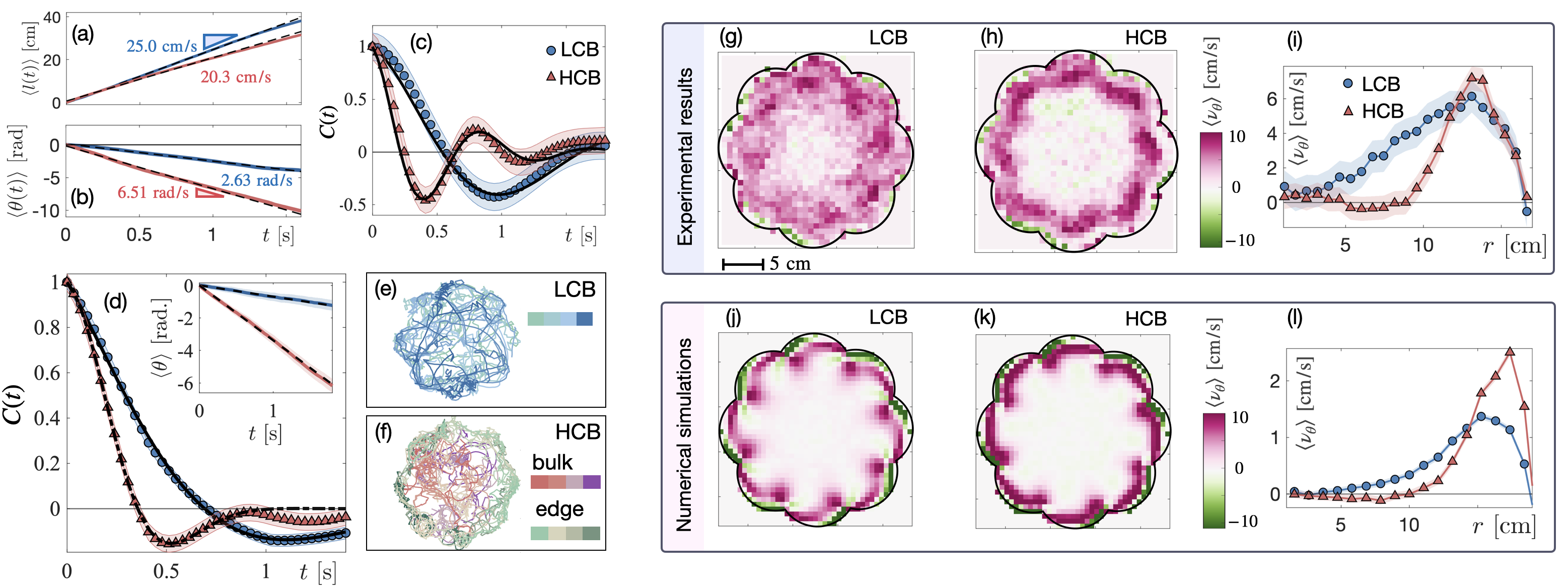}}
	\caption{\textbf{(a-c) Characterization of the individual bath particles properties.}
    (a) Ensemble averaged arc-length $l$. The bath particles self-propulsion speed $\nu_0$ is smaller for the modified bots forming the HCB.
    (b) Angular drift $\langle \theta(t) \rangle$ of the direction of motion of the bath particle both in the case of normal bots (blue) and in the case of modified asymmetric bots (red).
    (c) Autocorrelation function $C(t)$ of the direction of motion of the bath particles for normal bots (blue) and modified asymmetric bots (red). Both cases are fitted with the analytical expression given in the main text, which corresponds to chiral active Brownian motion including a Gaussian polydispersity of chirality, which is a free fitting parameter (black dashed lines).
    \textbf{(d-l) Collective dynamics of the active particles in the arena, global clockwise flows and their suppression.}
    (d) Correlation function $C(t)$ of the bots in the LCB (blue circles) and in the HCB (red triangles) within the arena. It is modified with respect to panel (c) by collisions between particles.
    $C(t)$ is fitted with the same expression as the free single bots autocorrelation functions shown in panel (c).
    In inset we show the linear mean drift of the direction of propulsion
    (e) Snapshots of trajectories in the LCB
    (f) Snapshots of trajectories in the HCB, the bulk trajectories (red to purple lines) are used to compute the autocorrelation function shown in panel (a), while the edge trajectories (shades of green) are not.
    (g) Global mean velocity field $\nu_\theta$ of bath particles along the azimuthal $\theta$ direction for the LCB composed of $20$ interacting unmodified bots. Pink color indicates a mean azimuthal clockwise flow while uncolored region denotes the absence of flows.
    The average is taken over the ensemble of trajectories of the bath particles and over the 20 minutes of recording.
    (h) Global velocity field $\nu_\theta$ for the HCB composed of 20 modified asymmetric bots.
    (i) Radial profile of the azimuthal velocity $\langle \nu_\theta(r) \rangle$, averaged over $\theta$ both for normal bots (blue circles) and modified bots (red triangles). The modified bath particles, with high chirality lead to a suppression of the flow in the center of the arena.
    (j, k, l) same results obtained from numerical simulations. 
    }
	\label{fig:BathChirality}
\end{figure*}

\subsection*{Properties of independent bots}

The dynamics of each type of active particles is quantitatively characterized in panels (a,b,c) of Fig.~\ref{fig:BathChirality}.
The self-propulsion speed $\nu_0 = dl/dt$, with $l = \int\sqrt{dx^2 + dy^2}$ the absolute length of the trajectory, is shown in Fig.~\ref{fig:BathChirality}~(a), slightly decreasing upon our design modification.
The chiral clockwise bias, measured via the mean angular drift of their self-propulsion direction $\omega = d\langle \theta(t)\rangle/dt$, is shown in Fig.~\ref{fig:BathChirality}~(b).
Our design modification leads to two-fold increase in chirality and a slight decrease if mean velocity $\nu_0$.
For the LCB-particles to $\omega = 2.63\pm 0.28~\rm{rad/s}$ and $\nu_0 = 25 ~\rm{cm/s}$ leading to the characteristic lenghscale $r_{\rm c} \equiv \nu_0/\omega = 9.5 ~\rm{cm}$ (a large radius corresponds to a low chirality).
For the HCB-particles to $6.51\pm0.33~\rm{rad/s}$ and $\nu_0 = 20.2 ~\rm{cm/s}$ hence $r_{\rm c} = 3.1~\rm{cm}$.

When measuring the autocorrelation function of orientation $C(t) = \langle \hat{n}(t)\hat{n}(0)\rangle$ in both cases (Fig.~\ref{fig:BathChirality}~(c)), we observe a damped oscillation, as expected for chiral active Brownian particles \cite{kummel2013circular}.
However, just like in biological chiral particles observed in nature, a key aspect is the polydispersity of the chirality $\omega$.
As shown in SI, including it leads to the correlation function $C(t) =\exp\left(-\frac{t}{\tau_p} - \frac{t^2 \varsigma^2}{2}\right)\cos(\omega t)$.
From the fit of this expression to the experimental data, we obtain the mean chirality $\omega = 2.92 \pm 0.23$ and $7.29 \pm 0.56~\rm{rad/s}$ for LCB and HCB respectively, the standard deviation of chirality $\varsigma$ and the persistence time $\tau_p$.
The fitted values of the mean chirality are close to the values obtained from the linear mean drift of the propulsion direction Fig.~\ref{fig:BathChirality}~(b).

\subsection*{Properties of the bots within the bath.}

When $N$ bots are confined in the arena, the finite area and the numerous collisions between bots modifies their effective properties.
In Fig.~\ref{fig:BathChirality}~(d), we show the orientation correlation function $C(t)$ of the bath particles within their many-body dynamics in the arena.
This shows that, despite collisions in the bath, the bots still approximately obey the dynamics of chiral active Brownian particles (with polydispersity in chirality, as discussed in Supplementary Information).
Their effective chirality is however reduced, as shown in the inset, with the mean angular drift of the orientation.
The fit of the correlation function shown in Fig.~\ref{fig:BathChirality}~(d) leads to an estimation of the correlation time of $\tau_p = 1.07 \pm 0.08~\rm s$ and $\tau_p = 0.58 \pm 0.08~\rm s$ for the LCB and HCB respectively as well as a the mean chirality $\omega = 2.34 \pm 0.08$  and $\omega = 4.29 \pm 2.7~\rm{rad/s}$ for LCB and HCB respectively.
These values differ from the chirality obtained by tracking the linear angular drift of the direction of motion (inset of Fig.~\ref{fig:BathChirality}~(d)) which leads to $\omega = 0.64 \pm 0.08$ and $2.37 \pm 0.1~\rm{rad/s}$ in LCB and HCB respectively.

To further characterize the collective behavior of the bots in the arena, we study their mean circulation.
In Fig.~\ref{fig:BathChirality}~(g, h), we show the average velocity $\nu_\theta$ along the azimuthal direction $\theta$ measured in our experiment.
A positive $\nu_\theta$, shown in pink denotes a clockwise velocity field.
Panel (a) reveals that the LCB leads to global clockwise flow, induced by the boundary and significant up to the center of the arena.
This is in line with flows observed with confined bacteria \cite{wioland2013confinement}.
In sharp contrast, panel (h) indicates that increasing single-particle chirality reduces bulk clockwise flow. Indeed, in the HCB, correlation length is shorter, and the boundary-driven currents remain confined to the edge of the arena \cite{li_robust_2024, kant2025edge}.
Increasing chirality thus insulates the bulk from the edge flows. 
Fig.~\ref{fig:BathChirality}~(i) quantifies this distinction via the radial distribution of the velocity average over the angle $\theta$. 
In the HCB case (red triangles), the bulk exhibits a $20$ \si{cm}-wide region in the center free from any mean current, i.e. it is effectively quiescent in the region where the measured odd response takes place (see Fig.~\ref{fig:Traj}).
This key control allows to rule out the hypothesis that the displacement along the $y$-axis is due to a mean flow.
This clear separation of boundary and bulk flows allows us to isolate tracer dynamics from background advection.
Minimal numerical simulations in a finite-size arena reproduce the same effect, as shown in Fig.~\ref{fig:BathChirality}(j, k, l).

As discussed in the main text, the tracer itself acquires an active chiral dynamics due to the interactions with the bath.
The correlation function of the tracer's direction of motion shown in Fig.~\ref{fig:BathChirality}~(e) also follows a decaying oscillatory profile.
Yet it obeys a distinct stretched exponential decay.
Its fit lead to an estimation of chirality of $2.64 \pm 1.14~\rm{rad/s}$ for the HCB, while no significant oscillatory chirality is measured from this fit in the LCB.
The tracer's chirality is also probed by the mean angular drift $\langle \theta(t)\rangle$ shown in Fig.~\ref{fig:BathChirality}~(f) that leads to an estimation of $\omega_{\rm T} = 0.14 \pm 0.08~\rm{rad/s}$ and $\omega_{\rm T} = 1.17 \pm 0.06~\rm{rad/s}$ for LCB and HCB respectively.

\begin{figure}[t!]
	\centerline{\includegraphics[width=0.85\linewidth]{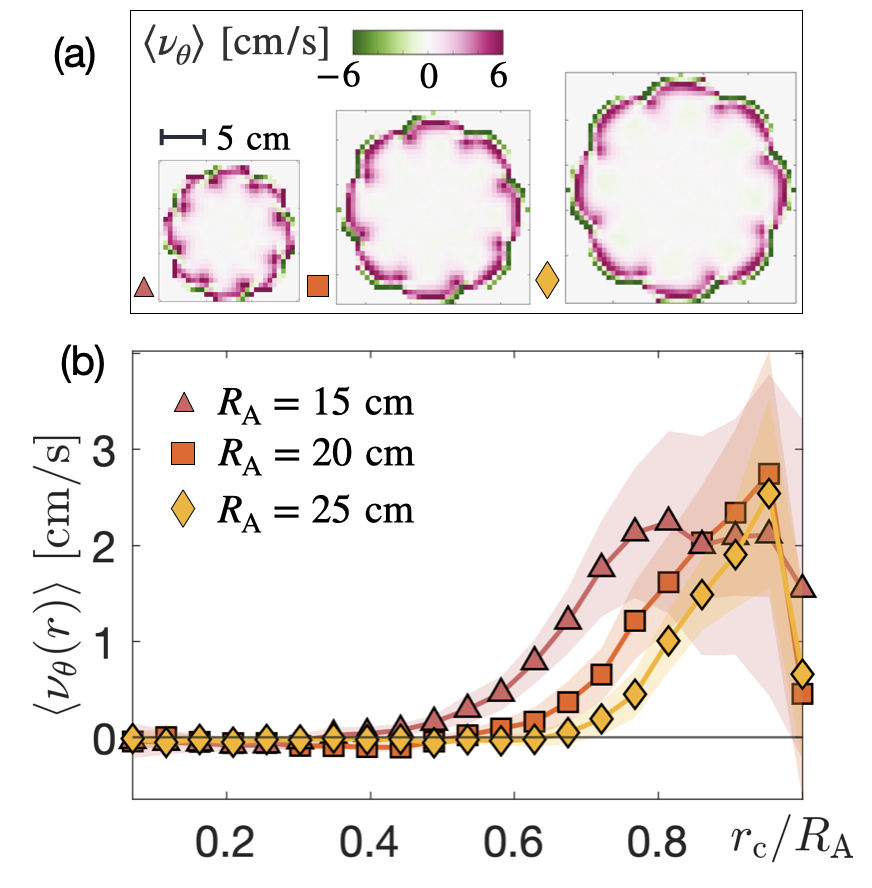}}
	\caption{\textbf{(a-d) Influence of arena length-scale}. (a) Bath azimuthal flow $\langle \nu_\theta\rangle$ shown as a color gradient for fixed bath particles parameters [$r_c = 3.1 ~\rm{cm}$; $\nu_0 = 20~\rm{cm/s}$] and three distinct arena radius $R_{\rm A} = 15 ;~20 ; $ and $25~\rm{cm}$. 
    (b) Average azimuthal currents $\langle \nu_\theta(r)\rangle$ as a function the distance from the center of the arena in the three configurations shown in panel (a). This plot reveals the progressive rejection of the currents to the edge of the arena.}
	\label{fig:ArenaSize}
\end{figure}

In Fig.~\ref{fig:ArenaSize}~(a, b), we show the azimuthal velocity field of the bath in the arena when its radius $R_{\rm A}$ is increased.
The clockwise edge currents are progressively rejected to the boundary of the arena as $r_c / R_{\rm A}$ becomes smaller.
This complements the experimental result presented Fig.~\ref{fig:BathChirality}~(g, h), indicating that chiral flows are suppressed in the bulk  as the ratio $r_c/R_{\rm A}$ increases.

\subsection*{Collective numerical simulations}


In this section we describe how we perform the numerical simulation presented in the main text.
To describe the motion of bots, the following two-dimensional underdamped $\dot{\mathbf{r}} = \bm{\nu}$ dynamics are used \cite{Caprini/etal:2024}:

\begin{eqnarray}
    m_{\rm bot}\frac{\dd \bm{\nu}_i}{\dd t} & + & \gamma\bm{\nu}_i \\
    & = &\gamma \nu_0 \mathbf{n}_i - \nabla\left[U^{\rm tot}_i + U_{\rm wall}(\mathbf{r}_i) \right] + \gamma \sqrt{2D}\boldsymbol{\xi}_i(t).\nonumber
\label{eq:dynamics}
\end{eqnarray}
Here, $\boldsymbol{\xi}_i(t)$ are Gaussian white noises with $\langle \boldsymbol{\xi}_i(t) \rangle = 0$ and $\langle \xi_{\alpha i}(t)\xi_{\beta j}(t')\rangle = \delta_{ij}\delta_{\alpha\beta}\delta(t-t')$; $\gamma$ is the friction coefficient, and $U^{\rm tot}_i= \sum\limits_{j \ne i}U_{\rm WCA}(|\mathbf{r}_i - \mathbf{r}_j|) + U_{\rm WCA}(|\mathbf{r}_i - \mathbf{R}|)$. In this way, we take into account only translation collisions between the bots as the main mechanism governing the tracer motion, and we do not account for rotational dynamics and rotational friction during the bot-tracer collisions as the angular dynamics is negligible for the passive tracer. Indeed, our simulations qualitatively capture the effects of tracer motion reported in the Results section. In order to neglect the impact of rotational dynamics on the translational motion, we set the upper limit of bots' and tracer's velocities to $\nu_0$ (i.e., the velocity of a single bot in the steady-state in the absence of noise).

The bots are modeled as ellipses with semi-major and semi-minor axes of lengths $a$ and $b$, respectively; with centers at positions $\mathbf{r}_i$ and orientations $\theta_i$ along their semi-major axis. The tracer is a disk with its center at position $\mathbf{R}$ and a radius $R_{\rm T}$. The arena is a circle with its center at zero and a radius $R_{\rm A}$.

The particles (bots and tracer) interact via the Weeks-Chandler-Anderson interaction, explicitly:

\begin{eqnarray}
	 & &U_{\rm WCA}(|\mathbf{r}_i - \mathbf{r}_j|)= \\
     &=& \begin{cases}
            4 \varepsilon \left[\left(\frac{\sigma_{ij}}{r_{ij}} \right)^{12} - \left(\frac{\sigma_{ij}}{r_{ij}} \right)^{6}\right] + \varepsilon, & \textrm{if} \ r_{ij} < 2^{1/6}\sigma_{ij}, \nonumber \\
            0, & \textrm{else}.
        \end{cases} 
\end{eqnarray}

Here, $r_{ij} = |\mathbf{r}_i - \mathbf{r}_j|$, and effective particle diameter $\sigma = \sigma_{ij}$ used in the simulations for ellipsoidal particles is calculated as \cite{Hahn:1999}:

\begin{eqnarray}
    & &\sigma_{ij} \\
    &=& \frac{2ab}{\left[\left(a^2 \sin^2\Theta_i + b^2\cos^2\Theta_i\right)\left(a^2 \sin^2\Theta_j + b^2\cos^2\Theta_j \right)\right]^{\frac{1}{4}}},\nonumber
\end{eqnarray}
where $\Theta_i$ is the angle between the ellipse orientation vector $\hat{\boldsymbol{\theta}}_i = (\cos\theta_i,\, \sin\theta_i)$ and vector $\mathbf{r}_i - \mathbf{r}_j$.

For bots-tracer interactions, $\sigma = \sigma_i$ is calculated as:
\begin{equation}
	\sigma_{i} = \frac{2\sqrt{ab R_{\rm T}}}{\left[\left(a^2 \sin^2\Theta_i + b^2\cos^2\Theta_i\right)\right]^{\frac{1}{4}}},
\end{equation}

The interaction with the arena walls ${{{{\bf{F}}}}}_{i}^{w}=-{{{{\bf{e}}}}}_{r}{\nabla }_{r}U_{\rm wall}(\mathbf{r})$ arises from the potential \cite{antonov2025self}:

\begin{equation}
    U_{\rm wall}(\mathbf{r}_i)  =\begin{cases}\displaystyle \frac{1}{2}\varepsilon_{\rm wall}{(r-R_s)}^{2},\quad &r\ge R_s,\\ 0,\hfill \quad &R_s > r\ge 0,\end{cases}.
\end{equation}

\begin{figure}[t!]
	\centerline{\includegraphics[width=0.8\linewidth]{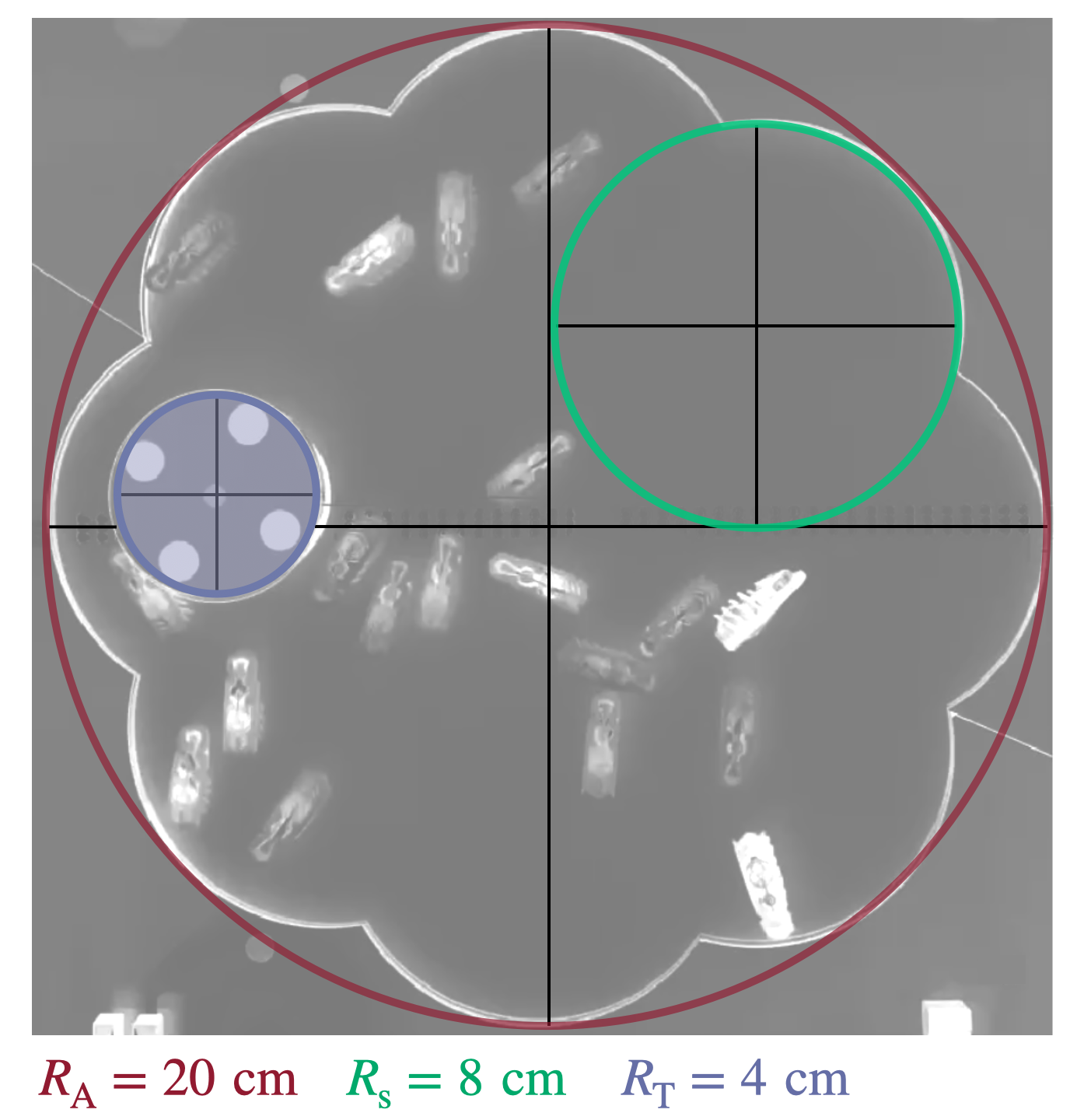}}
	\caption{\textbf{Geometry of the experimental setup}, showing the respective radius of the arena $R_{\rm A}$ (red circle), of each petal-like segment composing its boundary $R_{\rm s}$ (green circle) and of the tracer $R_{\rm T}$ (blue-gray circle).}
	\label{fig:Geometry}
\end{figure}

Here, $R_s$ denotes the radius of the segment ``petals'', and $r$ is the distance from the center of the corresponding circle to the furthest point of the bot or tracer (see Fig.~\ref{fig:Geometry} for the geometry of the experimental arena). In the simulations, the flower shape is constructed from eight circles of radius $R_s$, with their centers positioned at a distance of $R_A - R_s$ from the center of the arena.

The angular velocity in the overdamped regime is described as
\begin{equation}
	 \dot{\theta}_i = \omega + \kappa\sin\phi  + \sqrt{2D_{\rm r}}\eta_i.
\end{equation}
Here, $D_{\rm r}$ is the rotational diffusion, $\omega$ is chirality, $\eta$ is a Gaussian white noise, $\phi$ is the angle between the bot orientation $\hat{\boldsymbol{\theta}}$ and its velocity $\bm{\nu}$, and $\kappa$ is the self-alignment parameter setting the strength of the self-alignment torque \cite{Musacchio/etal:2025}. In our simulations, we use $\kappa=2.3$ (see the SI for the details) This torque aligns the orientation of a particle with its velocity and is typical for the bristle-bot systems due to asymmetric mass distribution combined with the leg-motor geometry \cite{Baconnier/etal:2022}.

The tracer dynamics is subjected to dry (Coulomb) friction and reads \cite{antonov2024inertial}:

\begin{table}[t!]
\begin{center}
\begin{tabular}[c]{||c || c || c ||}
	\hline
  Symbol & Meaning & Value \\
   \hline
   $m_{\rm bot}$ & Bristle-bot (bot) mass & 7.4 \si{\g} \\
   \hline
   $m_{\rm T}$ & Tracer mass & 13.4 \si{\g} \\
   \hline
   $m_{\rm ball}$ & Ball mass (Coulomb friction) & 0.87 \si{\g} \\
   \hline
   $m_{\rm ball}$ & Ball mass (Stokes friction) & $0.98 \cdot 10^{-1}$ \si{\g} \\
   \hline
   $R_{\rm A}$ & Arena radius & 20 \si{\cm} \\
   \hline
   $R_S$ & Arena's ``petal'' radius & 8.0 \si{\cm} \\
   \hline
   $R_{\rm T}$ & Tracer radius & 4 \si{\cm} \\
   \hline
   $2a$ & bot length & 4.3 \si{\cm} \\
   \hline
   $2b$ & bot width & 1.2 \si{\cm} \\
   \hline
   \specialcell{\vspace{-2ex} \\ $\displaystyle\frac{\varepsilon_{\rm part}}{m_{\rm bot}\nu_0^2}$ \vspace{1ex} \\ }  & bot interaction energy & $0.16-0.24$ \\
   \hline
   \specialcell{\vspace{-2ex} \\ $\displaystyle\frac{\varepsilon_{\rm wall}}{m_{\rm bot}\nu_0^2}$ \vspace{1ex} \\ } & Wall interaction energy & $1.6-2.4$ \\
   \hline
   $\nu_0$ & bot speed (HCB) & 25.01 \si{\cm/\sec} \\
   \hline
   $\nu_0$ & bot speed (LCB) & 20.24 \si{\cm/\sec} \\
   \hline
   $\omega$ & bot chirality (HCB) & 6.51 \si{rad/\sec} \\
   \hline
   $\omega$ & bot chirality (LCB) & 2.63 \si{rad/\sec} \\
   \hline
   $\mu$ & Dry friction coefficient & 0.48 \\
   \hline
   $\gamma$ & Stokes friction coefficient & $3.7 \cdot 10^2$ \si{\g/\sec} \\
   \hline
   $D$ & Tran.diffusion coefficient (HCB)& $0.1$ \si{\cm^2/\sec} \\
   \hline
   $D$ & Tran.diffusion coefficient (LCB)& $1$ \si{\cm^2/\sec} \\
   \hline
   \specialcell{\vspace{-2ex} \\ $D_{\rm r}\!=\!\displaystyle\frac{1}{\tau_p}$ \vspace{1ex} \\ } & Rot.diffusion coefficient (HCB) & $2$ \si{\sec^{-1}} \\
   \hline
   $D_{\rm r}$ & Rot.diffusion coefficient (LCB) & $1$ \si{\sec^{-1}} \\
   \hline
   $\kappa$ & Self-alignment strength & $2.3$ \\
   \hline
 \end{tabular}
\end{center}
\caption{Value of the parameters used in experiments and numerical simulations}
\label{Table1}
\end{table}

\begin{subequations}
\begin{equation}
	m_{\rm T}\frac{\dd \bm{\nu}}{\dd t} = -\nabla[U_{\rm tot} + U_{\rm wall}(\mathbf{R})] - \mu m_{\rm T} g \hat{\bm{\nu}} + m_{\rm ball} g\mathbf{e}_x.
\end{equation}
Here, $m_{\rm T}$ is the tracer mass, $\mathbf{e}_x$ is the unit vector along $x$-axis, $\bm{\nu}$ is its velocity, $\hat{\bm{\nu}}$ is the normalized vector which is defined to be zero when $\bm{\nu} = 0$, and $\mu$ is the dry friction coefficient. For a tracer subjected to Stokes friction instead, we take the dragging ball's mass such that the average velocities of the tracer along the drag directions are matched for both friction types considered. This procedure allows us to avoid overestimating the Hall effect due to geometry, i.e., for the resulting Hall angle, the displacement of the tracer along the drag force is identical for both types of friction. The resulting dynamics are given by
\begin{equation}
	m_{\rm T}\frac{\dd \bm{\nu}}{\dd t} = -\nabla[U_{\rm tot} + U_{\rm wall}(\mathbf{R})] - \gamma \bm{\nu} + m_{\rm ball} g\mathbf{e}_x.
\end{equation}
\end{subequations}

As bots exhibit dynamics close to the overdamped regime, we consider $\gamma$ high in the numerical simulations. The inertial term is necessary for a more accurate description of collisions between bots and the tracer.

\subsection*{Chiral active Ornstein-Uhlenbeck particle}

\begin{figure}[t!]
	\centerline{\includegraphics[width=0.95\linewidth]{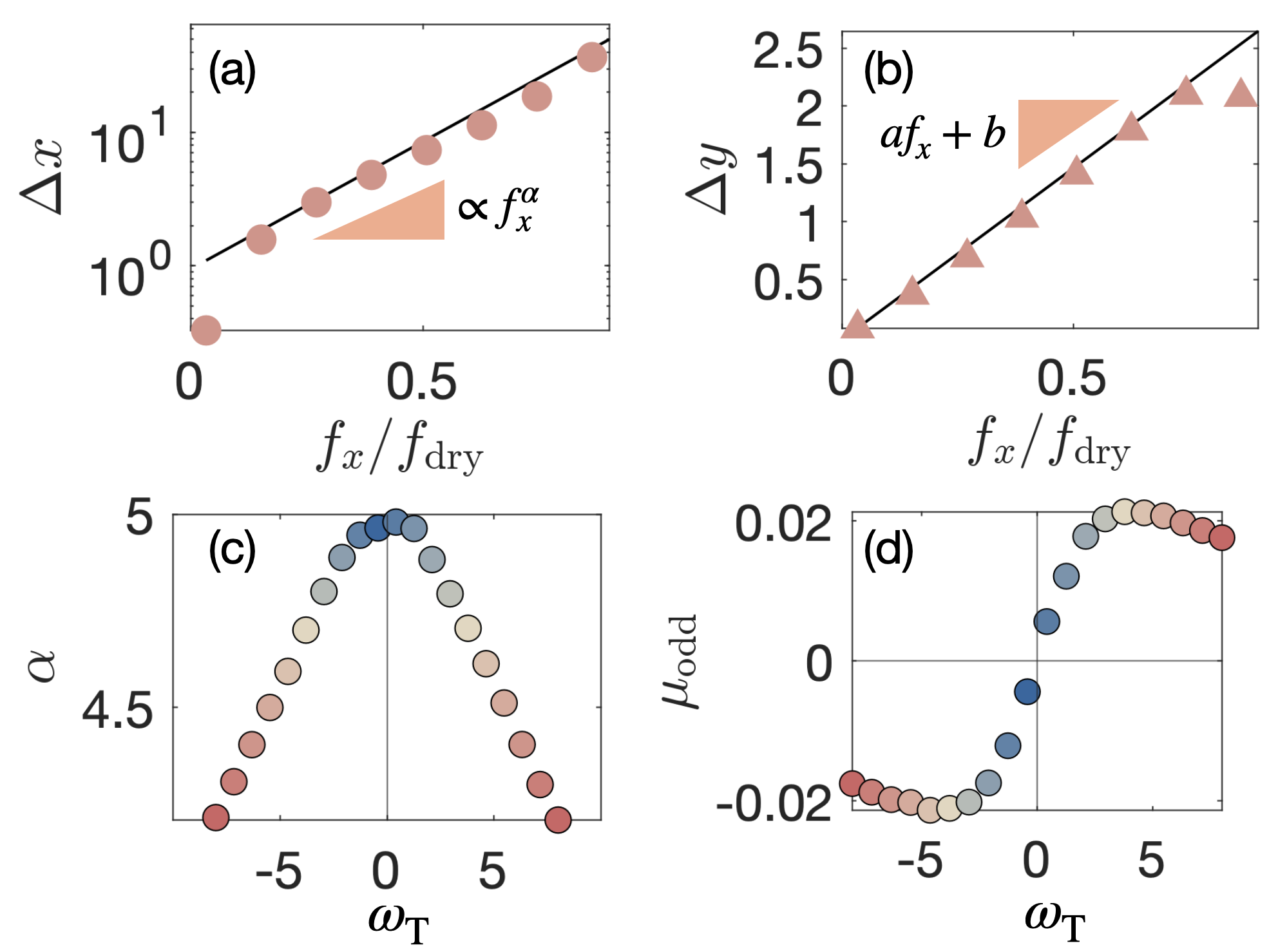}}
	\caption{\textbf{}
    (a) Mean displacement along the $x$-axis, averaged over an ensemble of 1000 trajectories simulated with the chiral AOUP model~\eqref{Eq:CAOUP} for parameters $\tau_p=0.25,\, f_A=2,\, f_x=1,\,f_{\rm dry} = 3$ using the Euler-Maruyama method with time step $dt=10^{-2}$ for total time $t=100$.
    (b) Mean displacement along the $y$-axis.
    (c) Exponent of the fitted power-law for displacement along the $x$-axis.
    (d) Slope of the linear fit of the odd displacement along the $y$-axis.}
	\label{fig:MobilityCoeff}
\end{figure}

As discussed in the main text, a chiral active Ornstein-Uhlenbeck particle (AOUP) experiencing dry friction forms a minimal model of the dynamics of the tracer.
In variance with Stokesian system \cite{poggioli_odd_2023}, we don't expect a linear relation between displacement and force.
The dynamics is a combination of stalls, ballistic movements and chiral loops.
In Fig.~\ref{fig:MobilityCoeff}, we explore the relation between the displacements $\Delta x$ (even) and $\Delta y$ (odd) with the applied force $f_x$.
Notably, the displacement along the $x$-axis follows a highly nonlinear trend, with exponents $\sim 4-5$ (see Fig.~\ref{fig:MobilityCoeff}(c)).
Yet the displacement along the $y$-axis are well described by a linear relation for small forces, while it decays towards zeros as $f_x$ approaches the dry friction threshold $f_{\rm dry}$.
Its slope gives the odd mobility coefficient $\mu_{\rm odd} = \nu_y / f_x$.

\subsection*{Note about the spinning degree of freedom}

A Hall effect is also observed in fluid systems without chiral bath constituents, for spinning objects (sometimes also referred to as the Magnus effect). Microscale experiments on the Magnus effect have shown that, similar to our case, going beyond instantaneous Stokes friction enhances odd transport \cite{cao2023memory}. However, the nature of the odd response in our system is different, since it arises from the bath chirality.
The tracer is experimentally barely spinning, and such tracer rotation is entirely neglected in simulations. In spite of this simplification, where the angular momentum transfer during collisions is ignored, the results remain qualitatively unchanged. This confirms that translational collisions between chiral bath particles and the passive tracer are sufficient to generate the observed odd response, and that tracer rotation plays no significant role.
This rules out that the odd mobility observed is due to the Magnus effect \cite{Kumar2019, cao2023memory}.
It agrees with the mechanism proposed in Fig.~\ref{fig:Chirality}(f) and in Fig.~\ref{fig:Traj}(a).

\bibliography{Chiral}

\section*{Supplementary Information}

\subsection*{Impact of self-alignment on the collisions between bristle-bots and tracer}

The self-alignment parameter $\kappa$ is determined from a controlled single-collision experiment between a single bot and a passive tracer. Experimentally, $\kappa$ is estimated from the condition that a bot propelled \textit{directly toward} the tracer undergoes a single, effectively elastic collision. This collision displaces the tracer and causes the bristle-bot (bot) to reorient before continuing its motion.

Under these conditions, the tracer’s displacement during a single impact can be computed using momentum- and energy-conservation relations:

\begin{eqnarray}
    m_{\rm bot} \nu_0 = m_{\rm bot} \nu_1 + m_{\rm T} \nu_{\rm coll};\\
    m_{\rm bot} \nu_0^2 = m_{\rm bot} \nu_1^2 + m_{\rm T} \nu_{\rm coll}^2,
\end{eqnarray}
where $\nu_1, \nu_{\rm coll}$ are the bot's and tracer's velocities after the collision, respectively; the latter reads:
\begin{equation}
    \displaystyle \nu_{\rm coll} = \frac{\nu_0}{\frac{m_{\rm T}}{m_{\rm bot}} + 1},
\end{equation}
with the total displacement of the passive tracer after a single collision $\Delta h = \nu_{\rm coll}^2/(2 \mu g)$. Matching this theoretical displacement to the experimentally measured one provides an estimate for the self-alignment coefficient $\kappa$. 

In our simulations, we position the bot at a distance of $3$ cm from the tracer and orient the bot directly toward it. We then measure the tracer’s displacement at $t=10$ sec - time sufficiently long for the bot to run away from the tracer after the initial impact, yet early enough to exclude any subsequent collisions between the bot and tracer.

The resulting tracer displacement, evaluated as a function of the self-alignment parameter $\kappa$, is compared to the experimentally measured shift $\Delta h$. The value of $\kappa$ is then inferred from the simulated-experimental correspondence that yields the closest agreement between the two displacements.

\subsection*{Autocorrelation with polydispersity}

In the autocorrelation of velocity, the dispersion of chirality $\omega$ over the ensemble of particles leads to a correction $\sim t^2$ to exponential decay. Indeed, the correlation function conditioned on a value of chirality takes the standard form $C(t|\Omega)= \exp(-t/\tau_p) \cos(\Omega t)$ where $\tau_p$ is the inverse of the angular diffusion coefficient $D_r$.
Assuming a Gaussian distribution of chiralities with variance $\varsigma^2$ and mean $\omega$, and using the fact that $\langle \cos (\Omega t)\rangle_\omega = \cos(\omega t) \exp(-t^2\varsigma^2/2)$, one arrives at a correlation function $C(t) = \langle C(t|\Omega) \rangle_\Omega=\exp\left(-\frac{t}{\tau_p} - \frac{t^2 \varsigma^2}{2}\right)\cos(\omega t)$.
Used in the main text and in Fig.~\ref{fig:BathChirality}(e).
Beyond the mean chirality $\omega$, the fitting procedure allows to evaluate its polydispersity standard deviation  [$\varsigma = 0.15$ and $0.29~\rm{rad/s}$; LCB and HCB respectively] and the persistence time [$\tau_p =  1.15 \pm 0.20$ and $\tau_p = 0.52 \pm 0.13~\rm s$; LCB and HCB respectively].
The error on $\omega$ is the combination of the fitting error and the fitted polydispersity $\varsigma$.

\subsection*{Sign of circulation currents}

\begin{figure}[t!]
    \centerline{\includegraphics[width=1\linewidth]{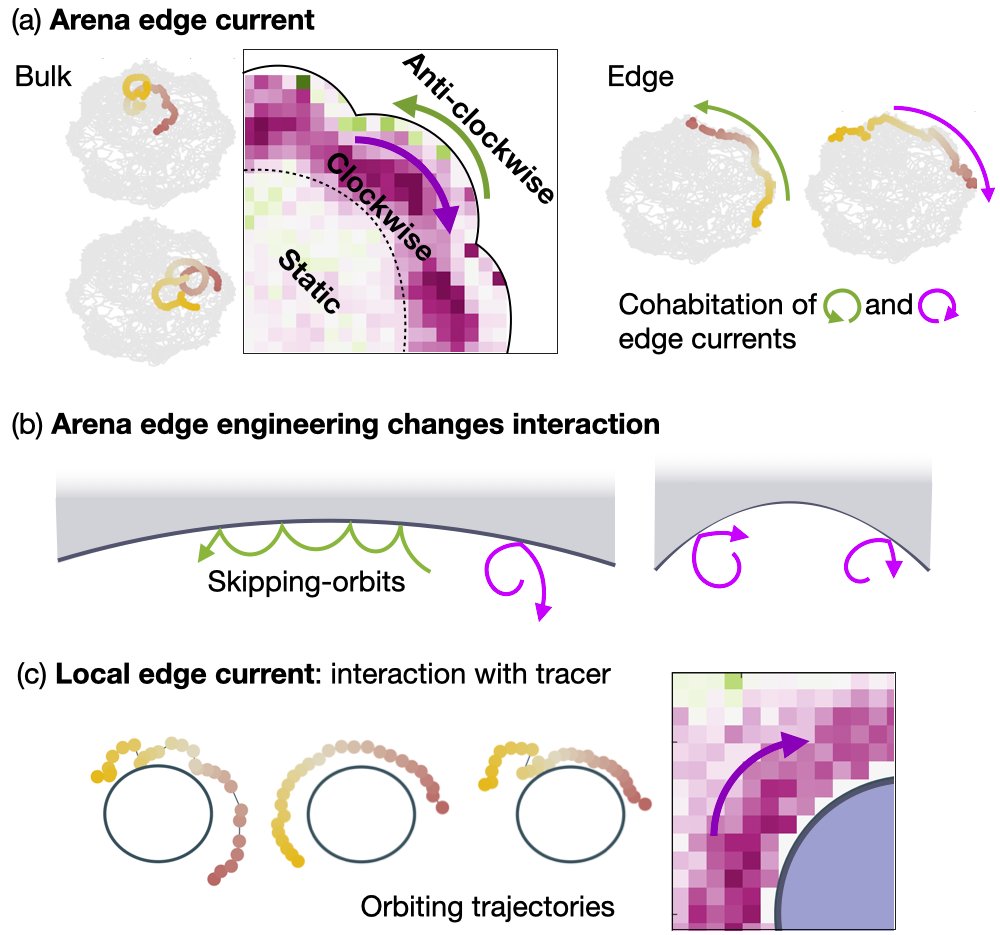}}
	\caption{\textbf{Interaction of the bots with the arena and the tracer}.
    (a) Structure of the edge currents localized near the arena's boundaries in the HCB case. The bulk fluid is composed of bots undergoing clockwise chiral trajectories (left) but is at rest on average. Near the edge, a clockwise circulation (pink) emerge, combination of clockwise and anti-clockwise bots trajectories (right).
    (b) The circular octogonal shape of the arena locally increases its radius of curvature, reducing the probability of skipping orbit-like anti-clockwise trajectories with respect to a circular arena.
    (c) the circulation around the tracer is clockwise, composed of orbiting bot trajectories.
    }
	\label{fig:Edge}
\end{figure}

The current of bath particles that are shown along an inner boundary (the tracer, see Fig.~\ref{fig:Chirality}~(c, d)) and along an outer boundary (the boundary of the arena, see Fig.~\ref{fig:BathChirality}~(g,h)) have equal sign, in contrast with chiral bacterial vortices \cite{li_robust_2024}.
To gain insight on this difference, in Fig.~\ref{fig:Edge} we detail the structure of the current at the edge of the arena and around the tracer.

The edge current along the boundary of the arena is composed of both clockwise (CW) and counter-clowkwise (CCW) trajectories (see right of Fig.~\ref{fig:Edge}). The CCW trajectories result from skipping-orbit interaction with the boundary.
They are strongly suppressed by the circular octogonal shape of the arena, which locally increases its radius of curvature (Fig.~\ref{fig:Edge}~(b)).
In Fig.~\ref{fig:Edge}~(c) we recall the current along an inner boundary, with characteristic orbital trajectories.
The sign of flows in our system is determined by the not well-separated length-scales characterizing our setup ($r_{\rm c} = 3-9$, $R_{\rm T} = 4$ and $R_{\rm A} = 20$ cm.
In contrast, for bacterial flows, there is a clear separation between the particle's correlation length and the edge's radius of curvature \cite{li_robust_2024}.

\end{document}